\documentstyle[prb,aps,graphics]{revtex}
\begin{document}

\title{Coherent multiple Andreev reflections and current resonances in
SNS junctions}

\author{\AA{}.~Ingerman\footnote{Email: ingerman@fy.chalmers.se},
  G.~Johansson, V.~S.~Shumeiko and G.~Wendin \\
  Department of Microelectronics and Nanoscience, Chalmers
  University of Technology and G\"oteborg University, S-412 96
  G\"oteborg, Sweden}
\date{\today}

\maketitle

\begin{abstract}
  We study coherent multiple Andreev reflections in quantum SNS
  junctions of finite length and arbitrary transparency. The presence
  of superconducting bound states in these junctions gives rise to
  great enhancement of the subgap current. The effect is most
  pronounced in low-transparency junctions, $D\ll1$, and in the
  interval of applied voltage $\Delta/2<eV<2\Delta$, where the
  amplitude of the current structures is proportional to the first
  power of the junction transparency $D$. The resonant current
  structures consist of steps and oscillations of the two-particle
  current and also of multiparticle resonance peaks. The positions of
  the two-particle current structures have pronounced temperature
  dependence which scales with $\Delta(T)$, while the positions of the
  multiparticle resonances have weak temperature dependence, being
  mostly determined by the junction geometry.  Despite the large
  resonant two-particle current, the excess current at large
  voltage is small and proportional to $ D^2$.\\
  Pacs: 74.50.+r, 74.80.Fp, 74.20.Fg, 73.23.Ad
\end{abstract}

\section{Introduction}

Transport properties of small conducting structures are strongly
influenced by size effects. Oscillation of magnetoresistance in thin
metallic films, and quantization of conductance in narrow wires and
point contacts are examples of such effects. Size effects in
superconducting tunneling have attracted attention since early
experiments by Tomasch \cite{tomasch.prl.65}. In these experiments,
oscillations of the tunnel conductance as a function of applied
voltage were found for tunneling from a superconductor to a thin
superconducting film of an NS proximity bilayer. The geometric
resonance nature of the effect was clearly indicated by the dependence
of the period of oscillations on the thickness of the superconducting
film. Similar conductance oscillations for tunneling into a normal
metal film of NS bilayers were reported by Rowell and McMillan
\cite{rowell.prl.66}. Later on an even more pronounced effect -- steps
on the current-voltage characteristics of SINS junctions at applied
subgap voltages, $eV<2\Delta$ -- was observed by
Rowell\cite{rowell.prl.73} (for a review see
Ref.~\onlinecite{wolf.85}). In addition to the dependence on the
thickness of the N-film, the period of the current steps also shows
temperature dependence which scales with the temperature dependence of
the superconducting gap $\Delta(T)$. The current steps occur at
applied subgap voltages, $eV<2\Delta$, and they are understood as
resonant features due to quasiparticle tunneling through
superconducting bound states existing in INS wells at energies lying
within the superconducting gap, $|E|<\Delta$, de~Gennes---Saint-James
levels\cite{degennes.pl.63}. 

Recently, properties of superconducting bound states have attracted new
attention in connection with observations of conductance anomalies
in mesoscopic NS structures. Resonant oscillations of the subgap
conductance in mesoscopic quasi-ballistic NS junctions have been
reported by Morpurgo et~al\cite{morpurgo.prl.97}. The zero bias conductance
peak, a related resonant phenomenon in diffusive NS junctions, was
discovered even earlier\cite{kastalsky.prl.91,wees.prl.92} and then
intensively studied both theoretically and experimentally (for a review
see Ref.~\onlinecite{lambert.jphyscondmat.98}).

In these recent studies of mesoscopic junctions, attention was switched
from the single-particle current through superconducting bound states to
the two-particle (Andreev) current. The traditional view of subgap
current transport in proximity NINS and SINS structures considers
single-particle tunneling into bound states in the normal region of the
INS well\cite{wolf.85,arnold.prb.78,gallaher.prb.80}, implicitly
assuming that the normal region of the INS well plays the role of
equilibrium reservoir. Such a model is appropriate for low-transmission
tunnel junctions with low tunneling rate compared to the inelastic
relaxation rate. However, transparent mesoscopic structures are in a 
different transport regime where the bound levels are well
decoupled from the superconducting reservoirs, and where injected
quasiparticles escape from the INS well via Andreev
reflection.\cite{blonder.prb.82,arnold.lowtempphys.85}
Resonant two-particle current in quantum NINS junctions has been
theoretically studied in Refs.\onlinecite{riedel.prb.93,chaudhuri.prb.95}. 
In SNS junctions, the quasiparticles
may undergo multiple Andreev reflections (MAR) before they escape into
the reservoirs.\cite{klapwijk.physicab.82}  In a number of recent
experiments with ballistic SNS devices fabricated with high mobility
2D electron gas (2DEG)
\cite{takayanagi.prb.95,chrestin.prb.97,kutchinsky.prl.97,bastian.prl.98}
the highly coherent MAR transport regime has been realized.

The purpose of this paper is to develop a theory of coherent multiple
Andreev reflections which will be applicable to 2DEG-based SNS
junctions.  In 2DEG devices the separation of the superconducting
electrodes $L$ is typically larger than 200 nm, which is of the same
order of magnitude as the superconducting coherence length,
$\xi_0=\hbar v_{F}/\Delta$ ($v_{F}$ is the Fermi velocity of the 2D
electrons), and superconducting bound states are formed well inside
the energy gap. The presence of bound states in the junctions of
finite length gives rise to resonances in the MAR transport, which
dramatically affects the subgap current.  Furthermore, it is possible
in 2DEG devices to use electrostatic gates to reach the quantum
transport regime with a small number of electron modes and variable
transmissivity.  A theory of coherent MAR has earlier been developed
for short superconducting
junctions,\cite{arnold.lowtempphys.87,bratus.prl.95,averin.prl.95,cuevas.prb.96,shumeiko.lowtempphys.97}
$L\ll \xi_0$, where superconducting bound states do not play any
significant role.\cite{note1} Such a theory is consistent with the
physical situation in atomic-size superconducting point contacts,
\cite{derpost.prl.94,scheer.prl.97,scheer.nature.98,ludoph.prb.00}
where quantization of conduction modes has turned out to be very
helpful for detailed comparison between theory and experiment.  The
purpose of our study is to investigate the interplay between
superconducting bound state resonances and coherent MAR in long SNS
junctions, $L\gtrsim\xi_0$, in the quantum transport regime.

In a number of publications, the coherent MAR approach has been applied to long 
SNS junctions\cite{gunsenheimer.prb.94,hurd.prb.96}.
However, these studies were restricted to fully transparent junctions where 
the bound states are strongly washed out and the resonances are
not pronounced (in fact, as we will show, at zero temperature the
current in such junctions does not show any structures). We will study
junctions with arbitrary transmissivity, $0<D<1$ and pay special
attention to the low-transparency limit, $D\ll 1$, where the resonance 
effects are most pronounced.

The paper is organized as follows. In Sec.~\ref{SNINSjunc} we
derive a 1D model for a gated ballistic 2DEG device with one
transport mode. In Sec.~\ref{details} we construct a scheme for
calculating MAR amplitudes in terms of wave propagation in energy
space. In Sec.~\ref{lowtransp}, single current resonances are
studied, and Sec.~\ref{interres} is devoted to a discussion of the
interplay between resonances in multi-particle currents. The
properties of the total subgap current is discussed in
Sec.~\ref{discussion}.

\section{1D model for quantum SNS junctions}
\label{SNINSjunc}

\begin{figure}[ht]
  \begin{center}
    \scalebox{0.5}{\includegraphics{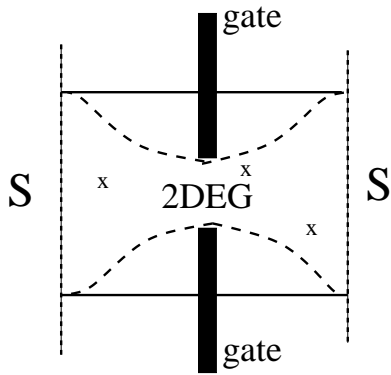}}
  \end{center}
  \caption{Sketch of the device: a ballistic 2DEG is sandwiched between
    two superconducting electrodes (S), and an electrostatic split
    gate creates a quantum constriction (dashed line) where only a few
    conducting modes are open; rare impurities are indicated with (x).}
  \label{S2DEGS-dev}
\end{figure}

We consider an SNS junction similar to the one discussed by Takayanagi
et~al.\cite{takayanagi.prb.95} schematically shown in
Fig.~\ref{S2DEGS-dev}. The junction consists of a normal conducting
channel fabricated with a high mobility 2DEG, which is confined
between superconducting electrodes. The distance between the
electrodes is comparable to the superconducting coherence length and
small compared to the elastic and inelastic mean free paths and to the
normal electron dephasing length. The superconductor-2DEG interfaces
are highly transmissive, the transmission coefficient typically
exceeding a value 0.75, and the number of conducting modes in the 2DEG
channel is controlled by a split gate.

Under these conditions, electrons ballistically move from one
electrode to the other while occasionally being scattered by rare
impurities or junction interfaces. Under a voltage bias applied to the
junction, the transport regime
corresponds to fully coherent multiple Andreev reflections (MAR).
To calculate the dc current we will apply the scattering theory approach
\cite{landauer.ibm.57,buttiker.prl.86,imry.86} generalized for
superconducting junctions; see
Refs.~\onlinecite{shumeiko.lowtempphys.97,ludoph.prb.00} and
references therein.

The normal electron propagation through the junction is generally
described by the $N$-channel scattering matrix. By assuming the split
gate to select only one transport mode, we will characterize the
transport through this mode by energy-dependent transmission
amplitude, $d(E)$, and reflection amplitudes, $r(E),r'(E)$ (the energy
$E$ is counted from the Fermi energy).  The scattering amplitudes
satisfying the unitarity relations, $dr^* + d^*r'=0$, $|d|^2+|r|^2
=1$.  The energy dispersion of the scattering amplitudes will
introduce the normal electron (Breit-Wigner) and superconducting
(Andreev) resonances in the scattering problem. The effect of narrow
Breit-Wigner resonances on coherent MAR was earlier studied by
Johansson et~al.\cite{johansson.physicac.97} and Levi~Yeyati
et~al.\cite{yeyati.prb.97} Here we will focus on the effect of Andreev
resonances and only consider Breit-Wigner resonances which are wide on
the scale of the energy gap. This will allow us to neglect the energy
dispersion of the junction transparency, $D=|d|^2\approx$const.
However, the scattering phases may depend on the energy, which yields
the Andreev resonances. In the simplest case, this dependence is a
linear function within the energy interval $|E| \sim\Delta$, and we
will write it on the form
\begin{equation}
  d(E)=d_0e^{iaE},\ r(E)=r_0e^{ibE},
\end{equation}
where $a,b$ are constant. In this case, the scattering properties
of the normal channel are similar to those of a 1D NIN junction.
Indeed, the corresponding 1D transfer matrix,
\begin{equation}
  \label{TE1}
\hat{T}(E)=\left(
\begin{array}{ll}
e^{-iaE}/d_0 & r_0^\ast e^{i(a-b)E}/d_0^\ast\\
r_0  e^{-i(a-b)E}/d_0 & e^{iaE}/d_0^\ast\\
\end{array}\right),
\end{equation}
can be decomposed into a product of three transfer matrices,
\begin{equation}
  \label{T}
  \hat{T}(E)= e^{-i\sigma_z L_{l}E/\Delta\xi_0} \hat{T}(0)
  e^{-i\sigma_z L_{r}E/\Delta\xi_0} \approx
  e^{-i\sigma_z k(E)L_{l}}\hat T e^{-i\sigma_z k(E)L_{r}},
\end{equation}
where $\sigma_{z}$ is a Pauli matrix. The first and the last
matrices describe ballistic propagation  of an electron, with wave
vector $k(E)=\sqrt{2m(E_F+E)}/\hbar\approx k_F+E/\Delta\xi_{0}$,
through the right and left N-regions of an effective junction with
lengths $L_{r}=b \Delta\xi_{0}/2$ and $L_{l}=(a-b/2) \Delta\xi_{0}
$ respectively (from right to left), and the matrix
$\hat{T}=e^{i\sigma_z k_FL_{l}}\hat T(0)e^{i\sigma_z k_FL_{r}}$
describes an effective barrier (I).

\begin{figure}[ht]
 \begin{center}
 \scalebox{0.4}{\includegraphics{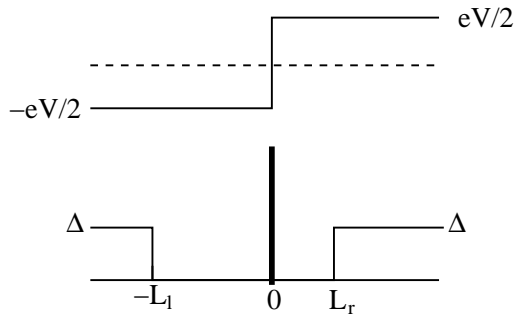}}
  \end{center}
  \caption{Spatial distribution of the superconducting order parameter
    and the electrostatic potential in the junction. The bold vertical line
    indicates the impurity potential. }
  \label{potentialdiag}
\end{figure}

Quasiparticle propagation through the effective 1D SNINS junction is
described
by means of the time-dependent Bogoliubov-de~Gennes (BdG) equation,
\cite{deGennes.book}
\begin{equation}
\label{BdG}
\left(\begin{array}{cc}
H_{0}(x) & \Delta(x) e^{\mbox{\small sgn}(x)ieVt/\hbar}\\
\Delta(x) e^{-\mbox{\small sgn}(x)ieVt/\hbar} & -H_{0}(x)
\end{array}\right)
\left(\begin{array}{c}
u(x,t)\\v(x,t)
\end{array}\right)
=i\hbar\partial_{t}\left(\begin{array}{c}
u(x,t)\\v(x,t)
\end{array}\right),
\end{equation}
where $H_{0}=\hat{p}^{2}/2m-\mu+ U(x) - \mbox{sgn}(x)eV/2$ is the
normal electron Hamiltonian, $U(x)$ is the impurity potential and $V$
is the applied voltage. The superconducting order parameter $\Delta(x)$
is constant within the superconducting electrodes and zero within the
normal region $-L_{l}<x<L_{r}$(see Fig.~\ref{potentialdiag}). In the
further calculations, the impurity potential is described by the
transfer matrix $\hat T$ in Eq.~(\ref{T}). The spatial distribution of
the applied potential along the channel is modeled with a step-like
function $\pm eV/2$. In fact, the actual spatial
distribution of the potential does not play any role in
this system: it can be included in the transfer matrix
in Eq.~(\ref{T}), leading to an additional energy-independent shift in
the scattering phases in the matrix $\hat T$. As we will see later
[comment after Eq. (\ref{resphase})], the energy-independent phases in
the $\hat T$-matrix do not affect the current, and can therefore be excluded.
The phase difference between the two superconductors follows from the
Josephson relation ($\dot{\phi}=2eV/\hbar$) and introduces time
dependence into the problem.  The superconducting electrodes are
considered to be equilibrium reservoirs where the quasiparticle wave
function is a superposition of electron- and hole-like plane
waves,
\begin{equation}
  \label{nambuvector}
  e^{\pm i\tilde{k}^{e}x-i(E \pm\sigma_{z}eV/2)t/\hbar}
  \begin{array}{cc}
    \left(\begin{array}{c}
        u\\v
      \end{array}\right), &
    e^{\pm i\tilde{k}^{h}x-i(E \pm\sigma_{z}eV/2)t/\hbar}
    \left(\begin{array}{c}
        v\\u
      \end{array}\right)
  \end{array}.
\end{equation}
In this equation, the $\pm$ signs in the time-dependent factors
refer to the left/right electrode, $\tilde{k}^{e,h}(E)$ is the
wave vector of electron/hole-like quasiparticles, and $u(E)$, $v(E)$
are the Bogoliubov amplitudes. The ratio of the Bogoliubov amplitudes
equals the amplitude of Andreev reflection for particles incoming from
the neighboring normal region,
\begin{equation}
\label{andreevdef}
\frac{v_{}}{u_{}} = a(E)=\left\{
\begin{array}{lr}
      \left(E-sgn(E)\sqrt{E^{2}-\Delta^{2}}\right)/
      \Delta,& |E|>\Delta \\
      \left(E-i\sqrt{\Delta^{2}-E^{2}}\right)/\Delta,&
      |E|<\Delta
    \end{array}
  \right..
\end{equation}
Since the time dependencies of the wave functions in the two
reservoirs are different, the quasiparticle scattering by the
junction is inelastic, and one has to consider a superposition of
plane waves with different energies in order to construct scattering states. 

\section{Calculation of current using scattering
states}
\label{details}
\subsection{Recursion relations for MAR amplitudes}

We will now proceed with the construction of recurrences for the
scattering amplitudes following the method suggested by Johansson et~al.
\cite{johansson.superlatt.99}. To this end we introduce
the wave functions in the left/right normal region
($l/r$) of the junction with respect to the position of the impurity. A
particular scattering state, labeled with the energy $E$ of the
incoming quasiparticle, will consist of a superposition of plane waves
with energies $E_{n}=E+neV$, where $n$ is an integer,
$-\infty<n<\infty$,
\begin{equation}
  \label{ansatz}
  \Psi_{l}(E)=\sum_{n=-\infty}^{\infty}
  e^{-i\left(E_{n}+\sigma_{z}eV/2\right)t/\hbar} \left[\left(
    \begin{array}{c}
      c_{n+}^{\uparrow,l}e^{ik^{e}_{n}x} +
      c_{n+}^{\downarrow,l}e^{-ik^{e}_{n}x} \\ 0\end{array}\right)
      +\left(\begin{array}{c} 0\\
      c_{n-}^{\uparrow,l}e^{ik^{h}_{n}x} +
      c_{n-}^{\downarrow,l}e^{-ik^{h}_{n}x}
    \end{array}
  \right) \right]
\end{equation}
\begin{displaymath}
  \Psi_{r}(E)= \sum_{n=-\infty}^{\infty}
  e^{-i\left(E_{n}-\sigma_{z}eV/2\right)t/\hbar} \left[\left(
    \begin{array}{c}
      c_{n-}^{\uparrow,r}e^{ik^{e}_{n}x} +
      c_{n-}^{\downarrow,r}e^{-ik^{e}_{n}x}\\
      0\end{array}\right)+
      \left(\begin{array}{c} 0 \\
      c_{n+}^{\uparrow,r}e^{ik^{h}_{n}x} +
      c_{n+}^{\downarrow,r}e^{-ik^{h}_{n}x}
    \end{array}
  \right)\right].
\end{displaymath}
The normal electron/hole wave vector $k_{n}^{e,h}$ is here defined as
$k_{n}^{e,h}=k(\pm E_{n})$, $k(E)=\sqrt{2m(E_{F}+E)}/\hbar\approx k_F+
E/\hbar v_F$. The meaning of the labels for the scattering (MAR)
amplitudes $c_n$ will be explained below.

Continuity of the scattering state wave function across the left and
right NS interfaces determines the relation between the electron and
hole amplitudes in the vicinity of each interface,
\begin{equation}\label{andrreflection}
c_{n+}^{\uparrow}= a_{n} c_{n-}^{\uparrow},\;\;\;
c_{n+}^{\downarrow} = a^{-1}_{n} c_{n-}^{\downarrow}, \;\;\; n\neq
0,\;\;\; a_{n}=a(E_{n}),
\end{equation}
which describes elastic Andreev reflection (l/r indices are omitted). It is convenient to consider scattering amplitudes near the impurity (at $x=\pm 0$) rather than at the NS interfaces, and to rewrite
Eq.~(\ref{andrreflection}) for
such amplitudes, combining the amplitudes of the ballistic propagation
through the normal regions with the Andreev reflection amplitude. Then,
in vector notation,
\begin{equation}
  \label{coeffdef}
  \hat{ c}_{n\pm}=\left(
  \begin{array}{c}
    c_{n\pm}^{\uparrow} \\
    c_{n\pm}^{\downarrow}
  \end{array}\right) ,
\end{equation}
the modified relation (\ref{andrreflection}) takes the form
\begin{equation}
  \label{Urelation}
  {\hat{ c}}_{n+}=\hat{U}_{n}{\hat{ c}}_{n-}, \;\;\;n\neq0,
\end{equation}
where
\begin{equation}\label{U}
  \hat{U}_{n}=e^{i\sigma_{z}E_n L_{l,r}/\Delta\xi_0}\left(
    \begin{array}{cc}
      a_{n} & 0 \\ 0& a_{n}^{-1}
    \end{array}
  \right)
e^{i\sigma_{z}E_n L_{l,r}/\Delta\xi_0}\equiv e^{i\sigma_{z}\varphi_{n}}.
\end{equation}
The phase
$\varphi_{n}=2E_{n}L_{l,r}/\Delta\xi_{0}-\arccos(E_{n}/\Delta)$,
characterizing $\hat{U}_{n}$, is real inside the energy gap,
$|E_{n}|\leq\Delta$, where it describes the total energy
dependent phase shift due to ballistic propagation and Andreev
reflection. Outside the gap, $|E_{n}|\geq\Delta$,
the phase $\varphi_{n}$ has an imaginary part which describes leakage
into the superconducting reservoirs due to incomplete Andreev
reflections.

By matching harmonics with the same time dependence in
Eq.~(\ref{ansatz}), we derive a relation between scattering
amplitudes at the left and the right side of the barrier:
\begin{equation}
  \label{Trelation}
  \hat{c}_{(n+1)-}^l=\hat{T}\hat{c}_{n+}^r,\;\;\;
\hat{c}_{(n+1)-}^r=\hat{T}^{-1}\hat{ c}_{n+}^l,
\end{equation}
where the effective barrier transfer matrix $\hat{T}$ is defined in
Eqs.~(\ref{TE1}), (\ref{T}).

\begin{figure}[ht]
  \begin{center}
   \scalebox{0.3}{\includegraphics{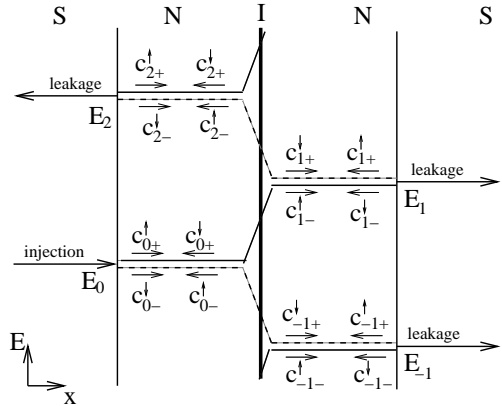}}
  \end{center}
  \caption{Scattering state in energy space: 
    coefficients $c_{n\pm}^{\uparrow\downarrow}$ correspond to the
    part of the scattering state at respective energy, location and
    specific direction. Electrons are indicated with full lines and
    holes with dashed lines; the arrows above (or below) each
    coefficient indicate the direction of the quasiparticle motion.  }
  \label{coeffpic}
\end{figure}

The recursion relations in Eqs.~(\ref{Urelation})
and~(\ref{Trelation}) couple the scattering amplitudes
$\hat{c}_{n\pm}$ into an infinitely large equation
system. This equation system describing coherent MAR is illustrated by
the MAR diagram in Fig.~\ref{coeffpic}.
The electron part of the quasiparticle injected at the left NS
interface propagates upwards along the energy axis, the amplitudes for
this propagation being labeled with $c^{\uparrow}$. At the injection
energy
$E=E_{0}$ (amplitude $c_{0+}^{\uparrow}$), the quasiparticle is
accelerated
across the barrier (I), where the potential drops. Thus, it enters the
right normal part of the junction with energy $E_1$
($c_{1-}^{\uparrow}$), undergoes Andreev reflection and goes back as a
hole ($c_{1+}^{\uparrow}$), entering the left normal part of the
junction having been accelerated to energy $E_2$
($c_{2-}^{\uparrow}$), and is then again converted into an electron
($c_{2+}^{\uparrow}$). The $\pm$
indices label the amplitudes after ($+$) and before ($-$) the Andreev reflection
for propagation upwards along the energy axis. There is a similar
trajectory of injected holes, which descends in energy, with the MAR
amplitudes labeled with $c^{\downarrow}$. Due to electron back
scattering at the barrier, the upward and downward propagating waves
are mixed, e.g. $c_{0+}^{\uparrow}$ being not only forward scattered
into $c_{1-}^{\uparrow}$, but also back-scattered into
$c_{0+}^{\downarrow}$, which opens up the possibility of interference.

Injection from the left reservoir, shown in Fig.~\ref{coeffpic},
generates a MAR path which
only connects even side bands at the left side of the junction with
odd side bands at the right side. Injection
from the right reservoir will generate a different MAR path, with even
side bands at the right side of the junction, i.e. the diagram in
Fig.~\ref{coeffpic} will effectively be mirrored around the
barrier (I). Thus, there are two independent equation
systems for the MAR amplitudes: injection from the left
and from right. The $l,r$ labels in the MAR amplitudes can then be 
omitted since they are uniquely defined by the source term and the side
band index.

The transport along the energy axis generated by MAR, from energy $E$ to
$E_n$, is conveniently described by
the effective transfer matrix $\hat{M}_{n0}$,
\begin{equation}
  \label{MAR}  \hat{ c}_{n-}=\hat{M}_{n0}\hat{ c}_{0+}, \ \ n>0
\;\;\;\hat{ c}_{n+}=[\hat{M}_{0n}]^{-1}\hat{ c}_{0-}, \ \ n<0
\end{equation}
\begin{equation}
  \label{mdef}
  \hat{M}_{nm}=\hat{T}_{n-1}\hat{U}_{n-1}\ldots
  \hat{U}_{m+1}\hat{T}_{m}, \;\;n>m,
\end{equation}
where $\hat{T}_{2k}=\hat{T}^{-1}$ and $\hat{T}_{2k+1}=\hat{T}$
for the injection from the left (for the injection from the right, the
even and odd side band indices are interchanged). For paths within the
superconducting gap, $|E_n,E_m|<\Delta$, the matrix $\hat{M}_{n0}$
satisfies the standard transfer matrix equation,
$\hat{M}_{nm}^{\phantom\dagger}\sigma_z\hat{M}_{nm}^{\dagger}=\sigma_z$,
which provides conservation of probability current along the energy
axis,
\begin{equation}
  \label{jp}
j^p_{n\pm}=
\hat{c}_{n\pm}^{\dagger}\sigma_z\hat{c}_{n\pm}^{\phantom\dagger} ,\;\;\;
j^p_{n\pm}= j^p_{m\pm},\;\;\; |E_{n,m}|<\Delta.
\end{equation}

An important consequence of the coherence of MAR is the
possibility of transmission resonances in energy space. From the form
of the $\hat{M}$-matrix,
\begin{equation}
  \label{mresstruc}
  \hat M_{n0}=\ldots \hat T^{-1}e^{i\sigma_{z}\varphi_{k}}\hat T\ldots,
\end{equation}
it is evident that when $\varphi_{k}=m\pi$, the two matrices
$\hat{T}^{-1}$ and $\hat{T}$ will cancel each other, and the
probability of transmission through this part will be unity, which
leads to resonant enhancement of MAR. The solutions $E^{(m)}$ of
the resonance equation,
\begin{equation}
  \label{resphase}
  \varphi^{(m)}_{k}=\varphi_{k}-
m\pi=\frac{2E_{k}L_{l,r}}{\Delta\xi_{0}}-
  \arccos {\frac{E_{k}}{\Delta}}-m\pi=0,
\end{equation}
coincide with the spectrum of the de~Gennes--Saint-James
levels localized in INS quantum wells \cite{degennes.pl.63}.
The corresponding bound states are located either on the left or the
right side of the junction.

Without loss of generality, the calculations can be performed with a
real matrix $\hat{T}$. The transformation to such a real matrix is
given by $\hat{T}\rightarrow \hat{V}_1 \hat{T} \hat{V}_2$, with
diagonal unitary matrices $\hat{V}_{1,2}$ whose elements are
constructed with the scattering phases, which are energy-independent.
It is clear from Eq.~(\ref{mresstruc}) that since these
energy-independent matrices commute with the matrices $\hat{U}_n$,
they cancel each other, and the matrix $\hat{M}_{n0}$ undergoes a
similar transformation. This will lead to an overall phase shift of
the scattering state which does not affect the current.

It is interesting to consider the special case of fully transparent
junctions, $D=1$, which has been studied in the literature
\cite{gunsenheimer.prb.94,hurd.prb.96}. In this case, all matrices
$\hat{T}_n$ in Eq.~(\ref{mdef}) are equal to the unity matrix, and the
$\hat{M}$-matrix takes the simple form
$\hat{M}_{n0}=\exp(i\sigma_z\sum_{m=1}^{n-1} \varphi_m)$. The length
of the junction then enters only through the phase of the MAR
amplitudes, which drops out of the side band current. Thus the dc
current of fully transparent SNS junctions is independent of length
and equal to the current in quantum constrictions
\cite{averin.prl.95,bratus.prb.97}.  In particular, at zero
temperature this current does not show any structures in the subgap
current. It is also worth mentioning that in this particular case of
fully transparent SNS junctions, the $\hat{M}$-matrix is diagonal and
therefore a closed set of recursive relations can be derived for the
MAR probabilities (not just for the MAR amplitudes, as in the general
case), equivalent to the equations for distribution functions derived
in the original paper by Klapwijk, Blonder and Tinkham
\cite{klapwijk.physicab.82}.

Equation~(\ref{MAR}) describes ``source-free'' propagation along the
MAR ladder. To complete the set of equations for the MAR amplitudes we
need to take into account quasi-particle injection, which introduces a
source term in Eq.~(\ref{MAR}). To this end, let us consider a
quasiparticle incoming from the left superconducting electrode with
energy $E$, having a wave function of the form
\begin{equation}
  \Psi_{l}^{S}(E)=
  e^{-iEt/\hbar-i\sigma_{z}eVt/2\hbar} \left[\delta_{\nu
        e}e^{i\tilde{k}^{e}x}
    \left(\begin{array}{c} u\\ v \end{array} \right)+
\delta_{\nu h}e^{-i\tilde{k}^{h}x}
\left(\begin{array}{c} v\\ u \end{array} \right)
  \right] .
\end{equation}
The two terms in this equation refer to electron- ($\nu=e$) and
hole-like ($\nu=h$) injected quasiparticles. We now include this wave
function into the continuity condition at the NS interface at energy
$E$, which gives us the following relation between the MAR amplitudes
$\hat{c}_{0+}$ and $\hat{c}_{0-}$,
\begin{equation}
  \label{source}
  \hat{ c}_{0+}=\hat{U}_{0}\hat{ c}_{0-}+\hat{Y}
\end{equation}
\begin{displaymath}
  \hat{Y}(E)=(u_{}^{2}-v_{}^{2})\left(
    \begin{array}{c}
      \delta_{\nu e}/u_{} \\ -\delta_{\nu h}
      e^{-2iEL_{l}/\Delta\xi_{0}}/v_{}
    \end{array}
  \right).
\end{displaymath}
For quasiparticles injected from the right, a similar equation holds
with the substitutions $e\rightarrow h$, and $L_{l}\rightarrow L_{r}$.
Equations~(\ref{MAR}) and (\ref{source}) give a complete set of
equations for the MAR amplitudes with the boundary conditions
$\hat{c}_{\pm\infty}=0$ at infinity.

\subsection{Calculation of MAR amplitudes}

A formal solution of Eqs.~(\ref{MAR}) and~(\ref{source}), which is
useful both for numerical calculations and analytical investigations,
can be
constructed by reducing this infinite set of recursion relations to a
finite set by representing the MAR process above $E_{n}$ and below
$E_{0}$ by boundary conditions involving reflection amplitudes $r_{n+}$ and $r_{0-}$, defined as
$c_{n+}^{\downarrow}=c_{n+}^{\uparrow}r_{n+}^{\phantom \uparrow}$ and
$c_{0-}^{\uparrow}=c_{0-}^{\downarrow}r_{0-}^{\phantom \downarrow}$.
This gives the following representation for the vectors in
Eq.~(\ref{coeffdef}),
\begin{equation}
  \label{nullvectors}
  \hat{ c}_{n+}= c_{n+}^{\uparrow}\left(
    \begin{array}{c}
      1 \\ r_{n+}
    \end{array}
  \right), \;\;\;
\hat{ c}_{0-}= c_{0-}^{\downarrow}\left(
    \begin{array}{c}
      r_{0-} \\ 1
    \end{array}
  \right).
\end{equation}
The reflection amplitudes $r_{n+}$ and $r_{0-}$ are independent of the
injection, in contrast to the coefficients $c_{n+}^{\uparrow},\
c_{0-}^{\downarrow}$. Furthermore, they are determined by the boundary
conditions $\hat{c}_{\pm\infty}=0$ and can be expressed in terms of the matrix elements of
$\hat{M}_{N n}$ and $\hat{M}_{0(-N)}$, where $N\rightarrow\infty$,
\begin{equation}
\label{lim}
 \lim_{N\rightarrow\infty}\hat M_{Nn} \left(
    \begin{array}{c}
      1 \\ r_{n+}
    \end{array}
  \right)=0,\;\;\;
\lim_{N\rightarrow\infty}[\hat M_{0(-N)}]^{-1}\left(
    \begin{array}{c}
      r_{0-} \\ 1
    \end{array}
  \right)=0.
\end{equation}
In other words, the vectors in Eq. (\ref{lim}) are equal to the
asymptotical values of the eigenvectors of
$\hat M$-matrices corresponding to the eigenvalues which decrease
when $N$ goes to infinity.
The advantage of introducing the reflection amplitudes $r_{n+}$ and
$r_{0-}$ is that although they have to be calculated numerically,
the recurrences which they obey do not contain resonances and converge
rather quickly.  This is in contrast to the matrix $\hat{M}_{n0}$,
which does possess resonances, but which can be calculated
analytically in a straight forward way for any given $n$.

The solutions of the recursion equations Eqs.~(\ref{Urelation})
and~(\ref{Trelation}) can now be explicitly written down. For any given energy $E$ we get four
different sets of solutions for four scattering states including
electron/hole injection from the left and the right.  Using the formal
expression in Eq.~(\ref{nullvectors}) and the
matrix elements of $M_{n0}= \left(\begin{array}{cc} m_{11} & m_{12} \\
m_{21} & m_{22} \end{array}\right)$, the solutions for injection
from the left ($n>0$) have the form,
\begin{equation}
  \label{coeffsol}
  \hat{ c}_{n+}=\frac{u_{}(1-a_{0}^{2})e^{i\varphi_{n}}
    \left[\delta_{\nu e}+a_{0}r_{0-}\delta_{\nu
        h}\right]} {m_{22}+m_{21}r_{0-}e^{2i\varphi_{0}}
    -m_{12}r_{n+}e^{2i\varphi_{n}}
    -m_{11}r_{0-}r_{n+}e^{2i\varphi_{0}+2i\varphi_{n}}}
  \left(\begin{array}{c}
      1\\ r_{n+}
    \end{array}\right).
\end{equation}
The solutions for injection from the right can be found through
interchanging
$e\leftrightarrow h$ and calculating all quantities with respect to
injection from the right. The solutions for $n<0$ are calculated in a
similar manner.

\subsection{Calculation of current}

Now turning our attention to the current, we calculate it in the
normal region next to the barrier, using the wave function in this
region, $\Psi_{}$, and assuming quasiparticle equilibrium within the
electrodes. The current then takes the form,
\begin{equation}
  I(t)=\frac{e}{hk_{F}}\int^{-\Delta}_{-\infty}dE\
  (u_{}^{2}-v_{}^{2})^{-1} \sum_{e/h,l/r}
  \mbox{Im}\left\{\Psi^{\dagger}_{}\frac{\partial}{\partial 
      x}\Psi^{\phantom \dagger}_{}\right\}\tanh{|E|\over 2k_BT}, 
\end{equation}
where $(u_{}^{2}-v_{}^{2})^{-1}=|E|/\sqrt{E^{2}-\Delta^{2}}=|E|/\xi$ is
the superconducting density of states, and the sum is over the four
scattering states at a given energy $E$ associated with the
electron- and hole-like quasiparticles (e/h) injected from the left and
right (l/r).
The current can be divided into parts with different time dependence
and expressed as a sum over harmonics:
\begin{equation}
  \label{Idef}
  I(t)=\sum_{N} I_{N}e^{2iNeVt/\hbar}
\end{equation}
Focusing on the dc ($N=0$) component, and calculating the contribution
of each scattering state at the injection side of the junction, we express the 
current spectral density $J(E)$ through the probability currents of electrons 
and holes at energies $E_{2n}$ (Fig.~\ref{coeffpic}),
$$ I_{dc}={e\over h}\int^{-\Delta}_{-\infty}dE\;J(E),$$
\begin{equation}
  \label{dccurrent}
  J(E)= \sum_{e/h,l/r} {|E|\over\xi}
  \sum_{n=-\infty}^{\infty}
  \left( \hat{ c}_{2n-}^{\dagger}
  \sigma_{z}\hat{ c}_{2n-}^{\phantom
  \dagger}+\hat{ c}_{2n+}^{\dagger}
  \sigma_{z}\hat{ c}_{2n+}^{\phantom \dagger}\right).
\end{equation}
These currents coinside with the probability currents $j^p_{n\pm}$, Eq.(\ref{jp}), flowing along the energy axis.

It is convenient to introduce a leakage current $J_{n}$, defined as the difference of the probability currents before and after Andreev reflection,
\begin{equation}
\label{leakage}
J_{n}=\sum_{e/h, l/r} {|E|\over\xi} \left (
 j^{p}_{n-}- j^{p}_{n+} \right).
\end{equation}
$J_{n}$ represents the amount of probability current from all the scattering
states injected at energy $E$ and leaking out
of the junction at energy $E_{n}$ (Fig.~\ref{coeffpic}). The leakage
current is zero inside the energy gap due to complete Andreev
reflection,  $J_{n}=0$, $|E_{n}|<\Delta$ [cf. Eq. (\ref{jp}].

The explicit expression for the leakage current for $n\neq0$ follows
from Eq.~(\ref{leakage}) after insertion of Eqs.~(\ref{coeffsol})
and~(\ref{Urelation}),
\begin{equation}
  \label{Ipn}
J_{n}=\sum_{l/r}
  \frac{(1-|a_{0}|^{2})(1-|a_{n}|^{2})(1+|r_{0-}a_{0}|^{2})
    (1+|r_{n+}a_{n}|^{2})} {\left|m_{22}+e^{2i\varphi_{0}} r_{0-
        }m_{21}-
      e^{2i\varphi_{n}}r_{n+}m_{12}-e^{2i\varphi_{0}}e^{2i\varphi_{n}}
      r_{0-}r_{n+}m_{11}\right|^{2}}.
\end{equation}
It follows from Eq.~(\ref{Ipn}) that the leakage currents are positive
for all $n\neq0$, $J_{n}\geq0$. One can also show that they
satisfy the inequality $\sum_{n\neq0}J_{n}\leq4$, which is a consequence of the conservation of probability
current: the leakage current of all side bands except of the side band
$n=0$ does not exceed the probability current injected into four
scattering states.  Furthermore, the leakage current satisfies the
important detailed balance equation \cite{johansson.superlatt.99},
\begin{equation}
\label{cancel}
J_{-n}(E)=J_{n}(E_{-n})
\end{equation}
i.e. the leakage at energy $E_{-n}$ due
to the injection at energy $E$ is the same as the leakage at energy
$E$ due to injection at energy $E_{-n}$.

Using the continuity of current across the barrier,
$j^{p}_{n+}=j^{p}_{(n+1)-}$, guaranteed by
the transfer matrix $\hat{T}$,
we can express the probability currents in Eq.~(\ref{dccurrent})
through the leakage current, Eq.~(\ref{leakage}),
\begin{equation}
\sum_{e/h, l/r} {|E|\over\xi}
j^{p}_{n-} = \sum_{k=n}^{\infty}
J_{k},\ n>0
\end{equation}
\begin{displaymath}
\sum_{e/h, l/r} {|E|\over\xi}j^{p}_{n+}
=-\sum_{k=-n}^{\infty}
J_{-k},\ n<0  ,
\end{displaymath}
by adding and subtracting consecutive terms in the sum. The spectral
density of the dc charge current Eq.~(\ref{dccurrent}) can then be
written
on the form
\begin{equation}
  \label{current_density}
J(E)= \sum_{n}n J_{n}(E),
\end{equation}
since $J_{n}$ appears in $n$ probability currents.
This formula has a clear physical meaning: the contribution to the
charge current of the $n$-th side band is proportional to the leakage
current of the side band times the effective transferred charge $ne$.

The detailed balance of the leakage currents, Eq. (\ref{cancel}), allows
us explicitly to prove that at zero temperature the scattering processes
between (occupied) states with negative energies, $E, E_{n}\leq-\Delta$ do not
contribute to the current, in agreement with the Pauli exclusion
principle. Indeed, by separating the contributions from
side bands with $n<0$ and remembering that the leakage current is zero
within
the gap, we get for zero temperature,
\begin{equation}\label{Ifinal}
I_{dc}=\frac{e}{h}\int_{-\infty}^{-\Delta}dE \sum_{n\neq0}n J_{n}(E)=
\sum_{n>0}\frac{ne}{h}\left(\int_{-\infty}^{-\Delta-eV}dE J_{n}(E)
+ \int_{\Delta-eV} ^{-\Delta} dE J_{n}(E)
- \int_{-\infty}^{-\Delta}dE J_{-n}(E)\right),
\end{equation}
where the first and the third terms cancel each other by virtue of Eq.
(\ref{cancel}). At finite temperature, these two terms produce current
of thermal excitations while  the second term gives the current of real
excitations created by the voltage source. Keeping only this term which
dominates at low temperature, we finally get,
\begin{equation}
\label{npartcurrent}
I_{dc}= \sum_{n>0}I_{n},\;\;\;
I_{n}= \frac{ne}{h}\theta(neV-2\Delta)
  \int_{\Delta-neV}^{-\Delta}dE
      J_{n}(E)\tanh(|E|/2k_BT).
\end{equation}
We end this section by noting a technically useful symmetry in the
current density, namely $J_{n}(E)=J_{n}(-E-neV)$, seen from
the explicit form of the $\hat{M}_{n0}$-matrix. This allows us to
reduce the integration interval in Eq.~(\ref{npartcurrent}) to
$-neV/2<E<-\Delta$.

\section{Current in terms of \MakeLowercase{$n$}-particle processes}
\label{lowtransp}

The approach formulated above provides necessary foundations for
numerical calculation of the current for arbitrary transparency and
length. However, to get a full understanding of the rich subgap
structure in the current-voltage characteristics, which may seem quite
random, especially for intermediate transparencies and lengths (see
Figs.~\ref{ItotD1plots}-\ref{ItotL4plots}), we will conduct a detailed
analytical study of the limit of low transparency $D\ll 1$. The
separation of currents into $n$-particle currents,
Eq.~(\ref{npartcurrent}), is our basis for
analysis, and we will study each current $I_{n}$ separately.

As explained in the previous section, the de~Gennes--Saint-James
levels, Eqs.~(\ref{mresstruc}) and~(\ref{resphase}), are important for
the
current transport through the junction, leading to resonant
enhancement of the current. Our main attention in this and the next
section is on the calculation of the position, height and width of the
main current peaks and oscillations which have the magnitude of order
$D$. To simplify notations, left/right injection indices are
omitted in most cases.

\subsection{Single-particle current}
\label{single}

\begin{figure}[ht]
  \begin{center}
   \scalebox{0.4}{\includegraphics{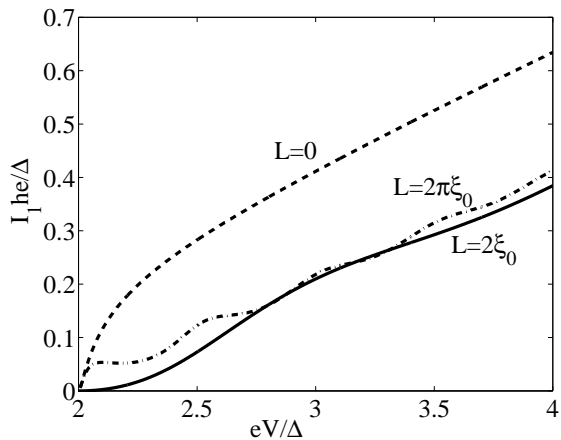}}
  \end{center}
  \caption{Single particle current for symmetric junctions
    $L_{l}=L_{r}=L/2$ for different junction lengths; the junction
    transparency is $D=0.1$. The current onset for the short junction
    ($L=0$) disappears for junctions with finite length (bold line);
    for $L=n\pi\xi_{0}$, the onset appears being roughly $n+1$ times
    smaller than the onset for $L=0$.}
  \label{I1D1plots}
\end{figure}

The single-particle current, which dominates at large applied voltages,
has, according to Eq.~(\ref{npartcurrent}), an onset at $eV=2\Delta$.
The full numerical solution for the single-particle current is
plotted in Fig.~\ref{I1D1plots}. The current shows pronounced
oscillations, and the magnitude of the slope at the current onset
strongly
depends on the junction length.

\begin{figure}[ht]
  \begin{center}
\scalebox{0.4}{\includegraphics{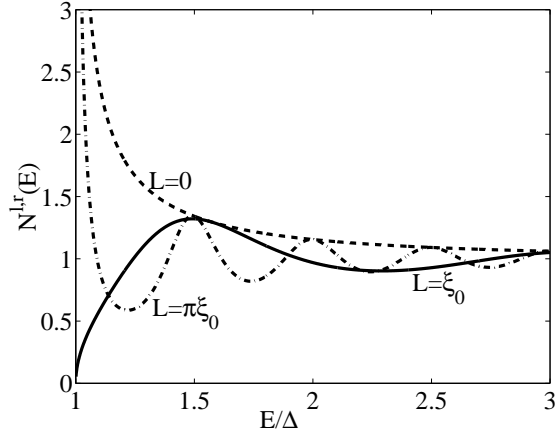}}
  \end{center}
  \caption{Density of states in the N region for different lengths $L$
    of the region. Singular behavior of the DOS for short junctions,
    $L=0$ (equal to the DOS in a superconductor), is suppressed for
    finite length junctions.  The amplitude of the first oscillation
    increases as the length increases, indicating accumulation of the
    spectral weight at the energy gap edge and formation of a bound
    state for $L=\pi\xi_{0}$. }
  \label{densplot}
\end{figure}

To understand this behavior, we analyze Eq.~(\ref{Ipn}) in the limit of
small transparency $D\ll1$, i.e. in the tunnel limit. First we note (see
appendix~\ref{refexp}) that the reflection amplitudes $r_{n+}$ and
$r_{0-}$ may be expanded as
\begin{equation}
  \label{reflectionexp}
  r_{n+}=(-1)^{n}\sqrt{R}+O(a_{n+1}^{2}D)
\end{equation}
\begin{displaymath}
  r_{0-}=\sqrt{R}+O(a_{-1}^{2}D),
\end{displaymath}
After inserting the explicit form of
$\hat{M}_{10}=\hat{T}$ together with the expansion~(\ref{reflectionexp})
into Eq.~(\ref{Ipn}) and putting $R=1$, we can write the single particle
current on the form
\begin{equation}
  \label{I1smallD}
  I_{1}=\frac{eD}{h}\theta(eV-2\Delta)\int_{\Delta-eV}^{-\Delta}dE
  \left[N^{l}(E_{})N^{r}(E_{1})+N^{r}(E_{})N^{l}(E_{1})\right]
\tanh(|E|/2k_BT),
\end{equation}
where
\begin{equation}
  \label{densityofstates}
  N^{l,r}(E)=\frac{|E|\sqrt{E^{2}-\Delta^{2}}}
  {E^{2}-\Delta^{2}+\Delta^{2}\sin^{2}
  \left(2EL_{l,r}/\Delta\xi_{0}\right)}.
\end{equation}
In analogy with the tunnel formula for the current, \cite{wolf.85}
$N^{l,r}$ is identified as the tunneling density of states (DOS) on
the left/right side of the junction. In Fig.~\ref{densplot} the
energy dependence of the DOS is presented. The deviation of this DOS
from the normal metal density of states is a manifestation of the
proximity effect. The expression~(\ref{densityofstates}) for the DOS has
earlier been derived for proximity NS sandwiches
\cite{wolfram.68,wolf.85,arnold.prb.78}. Note that the DOS in our case
is constant throughout the N-regions. In junctions with arbitrary
length, the DOS usually approaches zero at
the gap edge $|E|=\Delta$ (Ref.~\onlinecite{arnold.prb.78}).
Exceptions are junctions with lengths $L_{l,r}=m\pi\xi_{0}/2$ where a
bound state splits off from the gap edge.  In this case, the DOS
diverges at the gap edge.  The quantum well structure of the SNS
junctions also give rise to quasi-bound states in the continuum
spectrum, $|E|>\Delta$, seen as oscillations in the DOS.

The single particle current in Eq.~(\ref{I1smallD}) is written as the
integral over the product of the DOS at the entrance energy $E$ and
the exit energy $E+eV$. The latter depends on the applied voltage, as
well as the integration interval, and
therefore the DOS oscillations produce oscillations of the current
$I_{1}$ as a function of voltage (Rowell-McMillan oscillations
\cite{rowell.prl.66,rowell.prl.73}). The oscillations become
more pronounced when the junction is sufficiently long, and the
differential conductance may even become negative. It is also clear that
the DOS oscillates as a function of the length of the junction, which
give rise to oscillations also in $I_{1}$.

In short junctions, $L\ll\xi_0$, the current onset at $eV=2\Delta$ is
very steep, see Fig.~\ref{I1D1plots}. In junctions with finite length,
the current onset is smeared and replaced with a smooth oscillating
behavior. This can be directly related to the smearing of the
singularity in the DOS at the gap edge.  The length where the
crossover between these two behaviors occurs can be taken as a measure
of when finite length effects become important.  To estimate this
length, we write equation~(\ref{Ipn}) for small
lengths $L\ll\xi_{0}$, near the threshold, $eV=2\Delta+\Omega$,
$\Omega\ll\Delta$, keeping the first order terms in $D$ in the
denominator. For a symmetric junction, $L_{l}=L_{r}=L/2$, we get
\begin{equation}
  I_{1}=
  \frac{e\Delta\tanh(\Delta/k_BT)}{h}\int_{0}^{\pi}\frac{D\sin^{2}z\ dz}
  {\displaystyle \left(\sin z+ \frac{D\Delta}{4\Omega}\right)^{2}
  +\frac{L^{2}\Delta}{\xi_{0}^{2}\Omega}
  \left(1+\frac{D\Delta}{4\Omega}\right)}.
\end{equation}
From this formula it is clear that for short junctions ($L=0$),
the current onset has the width $\Omega\sim\Delta D/4$. If
$L$ is of the order of $\xi_{0}\sqrt{D}/2$, the size of the onset
has substantially diminished, and there is no visible onset at
$eV=2\Delta$ when $L\gg\xi_{0}\sqrt{D}/2$. This crossover between
steep onset and  smooth behavior, which happens already
for quite short lengths if $D$ is small, can be interpreted in
terms of a bound state which is situated exactly at the gap edge
in short junctions ($L=0$), and which moves down into the gap when
$L>0$, the effect becoming fully pronounced when the distance from
the gap edge, $\hbar v_F/L$, exceeds the dispersion of the Andreev
state, $\sqrt D\Delta$, in symmetric
junctions\cite{wendin.superlatt.96}.

When $L_{l,r}$ approaches $\pi\xi_{0}/2$, the lowest quasi-bound state
in
the continuum spectrum approaches the gap edge. This leads to an
accumulation of the spectral weight at the gap edge and reappearance
of the singularity in the DOS, which results in the reappearance of a
sharp current onset at $eV=2\Delta$, but with smaller magnitude; see
Fig.~\ref{I1D1plots} ($L_{l,r}=\pi\xi_{0}$).

It is of interest to note that in our calculations, based on the
scattering theory approach, the bound states are not directly involved
in the single-particle transport, which therefore is non-resonant and
shows no subgap resonance peaks. Within the tunnel model approach the
situation is qualitatively different: the DOS in Eq.~(\ref{I1smallD})
usually includes the contribution of the broadened bound states, and
therefore the single-particle current exists and has pronounced
resonant features at subgap voltages $eV<2\Delta$. This difference
results from the fact that, within the tunnel model approach, the
superconducting bound states are implicitly assumed to be connected to
the reservoirs (broadening due to inelastic interaction), which allows
a stationary current to flow through the bound states. In contrast,
within the scattering approach, the bound states are disconnected from
the reservoirs and have zero intrinsic width. In this case the bound
states obtain their width only due to higher order tunneling processes
involving Andreev reflections, which are manifested by the resonant
multi-particle currents. In practice, the relevance of the
multi-particle versus single-particle mechanism of the subgap current
transport is determined by physics and depends on the ratio of the 
corresponding dwelling and
relaxation times\cite{lofwander.supercond}. In this paper, the
inelastic relaxation time $\tau_i$ which determines the width of the
single-particle resonances is assumed to be much larger than the
dwelling time of the most important two-particle current,
$\tau_i\gg\hbar v_F/LD$.

\subsection{Two-particle current}
\label{twopart}

The two-particle current $I_{2}$ in quantum point contacts,
($L\ll\xi_{0}$) is of order $D^{2}$ when $eV<2\Delta$ and of order
$D^{2}\ln D$ when $eV>2\Delta$
(Ref.~\onlinecite{shumeiko.lowtempphys.97}). For finite length
junctions, the situation is different.  For the MAR paths where the
energy of the Andreev reflection coincides with a bound state, the
current spectral density $j^{p}_{2}$ is of order unity, due to
resonant transmission through this state. For low transparency
$D\ll1$, this gives a sharp concentration of the current density
around the resonant energies. In this limit, the two-particle current
is well described by the sum of contributions from these resonances,
and to evaluate them we examine the energy dependence of $J_{2}$
close to the resonant energies, $E_{1}=E^{(m)}+\delta E$.  Let us
consider the contribution to the leakage current $\left[J_{2}\right]^{l}$ from quasi-particles injected from the left. As shown in
appendix~\ref{twopartres}, in this case Eq.~(\ref{Ipn}) reduces to the
standard Breit-Wigner resonance form
\begin{equation}
\label{ip2res}
\left[J_{2}\right]^{l}=\frac{\Gamma_{0}^{(m)}
  \Gamma_{2}^{(m)}}{\displaystyle
  \left(\frac{\delta E-\delta E^{(m)}}{\Delta}\right)^{2}+
  \left(\frac{\Gamma_{0}^{(m)}+\Gamma_{2}^{(m)}}{2} \right)^{2}} ,
\end{equation}
where the tunneling rates $\Gamma_{n}^{(m)}$ are given by
$\Gamma_{n}^{(m)}=N^{l}(E_{n})D/2\eta^{(m)}$, $n=0,2$, and
\begin{equation}
\eta^{(m)}=\Delta\left.\frac{ \partial \varphi}
  {\partial
  E}\right|_{E=E^{(m)}}=\frac{2L_{r}}{\xi_{0}}+\frac{\Delta}
  {\sqrt{\Delta^{2}-(E^{(m)})^{2}}},
\end{equation}
and the position of the resonance is shifted by
\begin{equation}
  \delta E^{(m)} =\frac{D\Delta}{4\eta^{(m)}}\mbox{Im}
  \left\{\frac{1+e^{2i\varphi_{0}}}{1-e^{2i\varphi_{0}}}+
      \frac{1+e^{2i\varphi_{2}}}{1-e^{2i\varphi_{2}}} \right\}.
\end{equation}
An analogous result is valid for quasiparticles injected from the
right.

\begin{figure}[ht]
  \begin{center}
    \scalebox{0.4}{\includegraphics{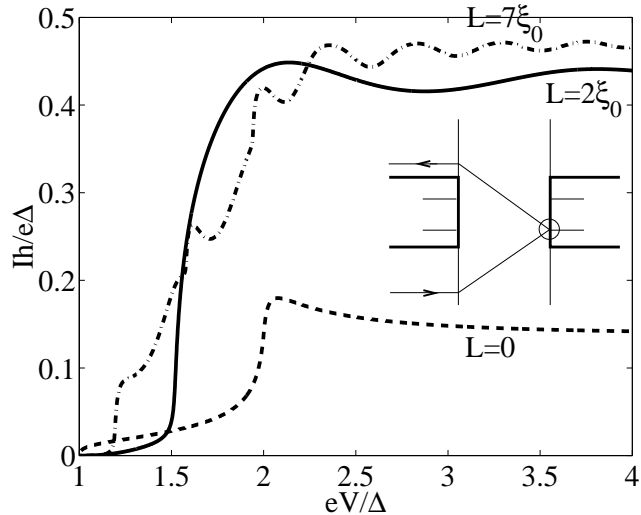}}
  \end{center}
  \caption{Two-particle current in symmetric junctions
    $L_{l}=L_{r}=L/2$ for different lengths; the junction transparency
    is $D=0.1$.  The resonant process shown in the inset becomes
    possible when $eV\geq\Delta+|E^{(m)}|$. }
  \label{I2D1plots}
\end{figure}

After integrating over energy, the two-particle current in the
resonance approximation may be written on the form
\begin{equation}
\label{i2approx}
I_{2}(eV) =\sum_{i=l,r}\sum_{m\geq0} \frac{2e}{h}
\theta\left(eV-E^{(m)}-\Delta\right)
\frac{2\pi D\Delta }{\eta^{(m)}}
\frac{N^{i}\left(E^{(m)}-eV\right)N^{i}\left(E^{(m)}+eV\right)}
{N^{i}\left(E^{(
m)}-eV\right)
+N^{i}\left(E^{(m)}+eV\right)}f^{(m)}(T,V),
\end{equation}
where the summation is over the positive bound level energies,
$0<E^{(m)}<\Delta$, and the DOS $N^{i}$ should be calculated at the
injection side of the junction and 
$f^{(m)}(T,V)=(1/2)\left[\tanh\left((eV-E^{(m)})/2k_BT\right)
  +\tanh\left((eV+E^{(m)})/ 2k_BT\right)\right]$. According to Eq. 
(\ref{i2approx}), the two-particle current $I_{2}(eV)$ increases in a
step-like manner in the voltage region $\Delta<eV<2\Delta$. The steps
occur at every voltage where a new resonant channel through a bound
state opens up, at $eV^{(m)}=\Delta+E^{(m)}$.  We note that the step
positions depend on temperature and approximately scale with
$\Delta(T)$. Each current step has the height of order $D$. As seen
from Eq.  (\ref{i2approx}), the contribution to the current of a
particular bound state $E^{(m)}$ is modulated, as a function of
voltage, by the oscillations of the density of states at the entrance
and exit energies, $N(E^{(m)}\pm eV)$. In other words, the pronounced
oscillations of the two-particle current seen in Fig.~\ref{I2D1plots}
reflect how close the entrance and exit energies $E^{(m)}\pm eV$ are
to a quasi-bound state in the continuum.  For $eV>2\Delta$, the
two-particle current $I_{2}$ oscillates around a constant value with
an amplitude of oscillation decreasing as $\Delta^{2}/eV^{2}$ for
large voltages.

It is interesting to compare the resonant structures of the two-particle
current with the resonant structures in NINS
junctions.\cite{riedel.prb.93,chaudhuri.prb.95}
In NINS junctions, the resonant current steps occur at $eV=E^{(m)}$,
and they do not have any modulation because the DOS on the normal
side of the junction is constant.

\begin{figure}[ht]
  \begin{center}
    \scalebox{0.4}{\includegraphics{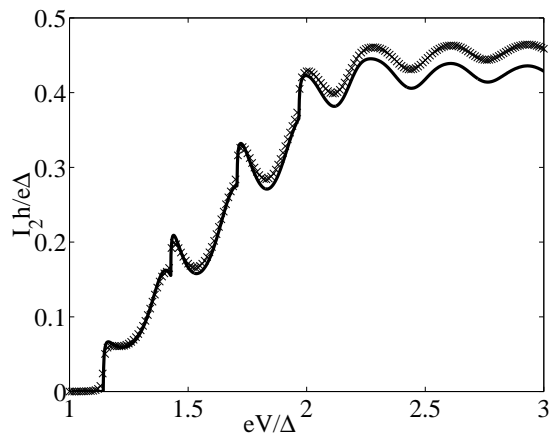}}
  \end{center}
  \caption{Comparison between the approximate expression for
    the two-particle-current in Eq.~(\ref{i2approx}) (solid curve)
    and the full numerical solution (crosses):
    $D=0.1$, $L_{l}=L_{r}=L/2=5\xi_{0}$. }
  \label{i2comp}
\end{figure}

The distance between the resonances and the resonance widths are
proportional to the bound level spacing, and they decrease in long
junctions. For sufficiently long junctions, the two-particle current may
thus give the appearance to include a series of peaks, as shown on
Fig.~\ref{i2comp}.
In symmetric junctions, the bound state energies at both sides of the
barrier will coincide, reducing the number of steps with a factor of
two, and giving current steps of double height.

We will conclude this subsection by noting that the difference between
the full numerical calculation of the two-particle current and the
resonant approximation given in Eq.~(\ref{i2approx})
is rather small already when $D=0.1$, as can be seen in
Fig.~\ref{i2comp}.

\subsection{Excess current}
\label{excesscurrent}

Excess current in SNS junctions, i.e. the difference between the current in
superconducting junction and in the normal junction at large voltage,
\begin{equation}
I^{exc}= I-G_{N}V+O(\Delta/eV),
\end{equation}
is commonly considered as a measure of the intensity of Andreev reflection.
In tunnel SIS junctions and low-transmissive point contacts the
excess current is small,
$I^{exc}\approx D^2 e\Delta/\pi\hbar$, $D\ll 1$, while in fully transparent contacts the excess current is large, $I^{exc} =
8e\Delta/3\pi\hbar$, $D=1$.\cite{shumeiko.lowtempphys.97} Accordingly,
one would expect large excess current in long SNS junctions due to the
resonant enhancement of the two-particle current.  However, the excess current is small because of a large deficiency, of order $D$, of the
single-particle current caused by the broadening of the current onset at
the threshold. As we will show, the single-particle and two-particle
currents undergo a fine cancellation, yielding small net excess current
of order $D^{2}$ when $D\ll 1$.

\begin{figure}[ht]
  \begin{center}
    \scalebox{0.4}{\includegraphics{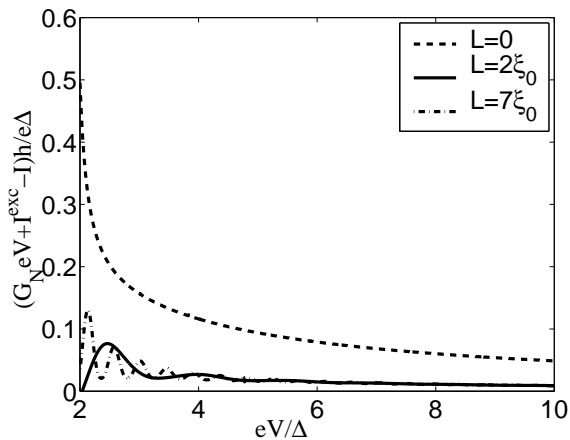}}
  \end{center}
  \caption{Deviation of the current from its asymptotical value at $V=\infty$:
    the excess current value is approached much faster in finite
    length junctions, here shown for $D=0.3$.}
  \label{excessapproach}
\end{figure}

\begin{figure}[ht]
  \begin{center}
    \scalebox{0.4}{\includegraphics{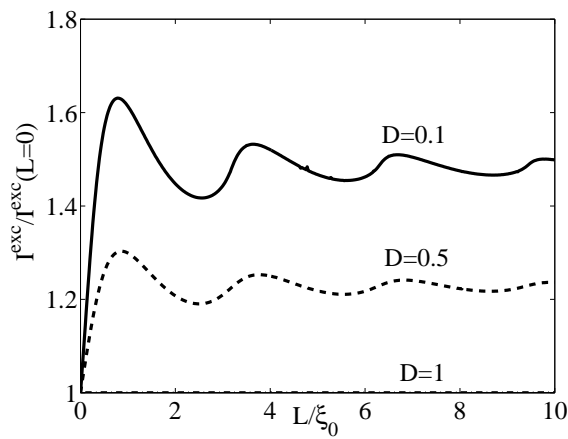}}
  \end{center}
  \caption{Dependence of the (normalized) excess current on the
    junction length for different transparencies. For fully
    transparent junction, $D=1$, the excess currents are identical for
    all junction lengths; the excess current increases for
    small-transparency junctions.  }
  \label{excessplot}
\end{figure}

The excess current has contributions only from
the single- and two-particle currents, since all higher order currents
include at least one Andreev reflection outside the gap whose
probability is of order $(\Delta/eV)^2$. In the limit of large
voltage, $eV\gg\Delta$, the relevant part of the current in
Eq.~(\ref{npartcurrent}) then takes the form
\begin{equation}
\label{Ihighvoltage}
I_{1} =\frac{4De}{h}\int_{-eV/2}^{-\Delta} dE\;
\frac{(1-a_{0}^{2})(1+Ra_{0}^{2})}
{1+R^{2}a_{0}^{4}-2R\ \mbox{Re}
\left\{e^{2i\varphi_{0}}\right\}}
\end{equation}
\begin{displaymath}
I_{2}=\frac{8D^{2}e}{h}\int_{-eV}^{-\Delta} dE\;
\frac{|a_{1}|^{2}}{1+R^{2}|a_{1}|^{4}-2R\ \mbox{Re}
\left\{e^{2i\varphi_{1}}\right\} }.
\end{displaymath}
These equations are written for symmetric junctions, $L_{r}=L_{l}=L/2$,
and for zero temperature; small Andreev reflection amplitudes
$\leq|a(eV/2)|\ll1$ have been
neglected in Eq.~(\ref{Ipn}). The behavior of the current in
Eq.~(\ref{Ihighvoltage}) as a function of voltage is presented in
Fig.~\ref{excessapproach} for different lengths. It is clearly seen
that the limiting value of the excess current is approached much
faster in finite length SNS junctions compared to point contacts
($L=0$). In Fig.~\ref{excessplot} the excess current behavior with
respect to the junction length is presented for different
transparencies.
To analytically examine the excess current in the limit of small
transparency, $D\ll 1$, it is convenient to start with equations
(\ref{I1smallD}) and (\ref{i2approx}).
To first order of $D$ the excess current assumes the form ($T=0$),
\begin{equation}
  \label{IexcsmallD}
  I^{exc}=I^{exc}_{1}+I^{exc}_{2},
\end{equation}
\begin{displaymath}
  I_{1}^{exc}=-\frac{4eD\Delta}{h}
  +\frac{2eD}{h}\int_{\Delta}^{\infty}\left[ N^{l}(E)+N^{r}(E)
  -2\right]dE,
\end{displaymath}
\begin{displaymath}
  I_{2}^{exc}=\sum_{l/r,\ m\geq0} \frac{2\pi D
    e\Delta}{h\eta^{(m)}}.
\end{displaymath}
Let us consider the contributions to the
single-particle current from the left electrode,
\begin{equation}
\label{I1excl}
\left[I^{exc}_{1}\right]^{l}=-\frac{2eD\Delta}{h}
+\frac{2eD}{h}\int_{\Delta}^{\infty}\left[ N^{l}(E)-1\right]dE.
\end{equation}
Inserting $N^{l}(E)$ from Eq.~(\ref{densityofstates}), this equation
can be transformed to the form,
\begin{equation}
\left[I^{exc}_{1}\right]^{l}=
\frac{2eD}{h}\int_{\Delta}^{\infty}\left(\frac{E\xi}
{\xi^{2}+\Delta^{2}\sin^{2}(2EL_{l}/\Delta\xi_{0})}-
\frac{E}{\xi}\right)dE=
-\frac{eD}{h}\int_{-\infty}^{\infty}d\xi
\frac{\sin^{2}(2EL_{l}/\Delta\xi_{0})}{\xi^{2}+\sin^{2}
(2EL_{l}/\Delta\xi_{0})},
\end{equation}
where $\xi=\sqrt{E^{2}-\Delta^{2}}$.
It is now possible to analytically continue the integral in the upper
half plane which will reduce the integral to a sum over the residues of
the poles given by the equation
$\xi^{2}+\sin^{2}(2EL_{l}/\Delta\xi_{0})$. Comparing this equation
with Eq.~(\ref{resphase}) we find that the poles
coincide with the energies of the bound states in the gap. The
excess current contribution from the left-injected single particle
current is thus
\begin{equation}
\left[I^{exc}_{1}\right]^{l}  = -\frac{2D\pi
e\Delta}{h}\sum_{m\geq0}
\frac{1}{\eta^{(m)}}=-\left[I^{exc}_{2}\right]^{l},
\end{equation}
where $\left[I^{exc}_{2}\right]^{l}$ is the contribution to the
two-particle current from the bound state resonances at the left
electrode. A similar relation is derived for current from the right
electrode. Thus, there is exact cancellation of the excess
single-particle and two-particle currents to first order in $D$.

It is interesting to note that the cancellation effect is related to
the conservation of the number of states in a proximity normal metal
compared to the conventional normal metal. It follows from
Eq.~(\ref{I1excl}) that $I_{1}h/2eD\Delta$ is equal to the difference
between the number of continuum states in the proximity metal and the
total number of states in a conventional metal, while, on the other
hand, the number of the bound states, is equal to
\begin{equation}
\int_{0}^{\Delta}
dE\ \sum_{m\geq0}\delta(\varphi(E)-m\pi)=\int_{0}^{\Delta}dE\
\sum_{m\geq0}\delta(E-E^{(m)})/\eta^{(m)}=
I_{2}h/2eD\Delta,
\end{equation}
according to Eq.~(\ref{IexcsmallD}).

\section{Interplay between resonances}
\label{interres}

For processes with several Andreev reflections ($n\geq3$), the
possibilities for resonances increase. Every Andreev reflection energy
may coincide with a bound state energy and thus be resonant. For some
specific voltages, more than one resonance is important, creating a
situation of overlapping resonances, which can enhance the current
giving peaks in the current-voltage characteristics at these voltages.

\subsection{Three-particle current}

The three-particle current $I_{3}$ has a non-resonant value of order
$D^{3}$.  However, $I_{3}$ is enhanced to order $D^{2}$ when the
energy of one of the two Andreev reflections coincides with a
bound state energy.
\begin{figure}[ht]
  \begin{center}
    \scalebox{0.4}{\includegraphics{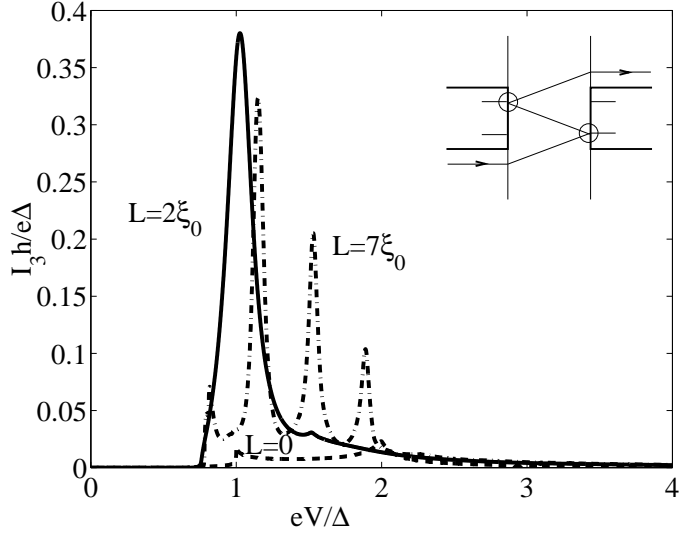}}
  \end{center}
  \caption{Three-particle current in symmetric junctions
    $L_{l}=L_{r}=L/2$ for different lengths; the junction transparency
is $D=0.1$. The MAR path with two overlapping resonances, shown on the inset, 
generates a current peak with height proportional to $D$.}
\label{I3D1plots}
\end{figure}
For the applied voltage equal to the difference between two
bound state energies, $eV^{(km)}=E^{(m)}-E^{(k)}$, two resonances occur
simultaneously, i.e. form a
resonance consisting of two overlapping single resonances; see the
inset in Fig.~\ref{I3D1plots}. This will enhance the current to order
$D$ close to this voltage, giving a peak in the IVC. The number of
peaks is equal to the number of bound state pairs. The
peaks are located in the voltage interval $2\Delta/3<eV<2\Delta$; we note that the peak positions are weakly dependent on temperature.

To evaluate the height and the width of these peaks, we study the
contribution from overlapping resonances at $E_{1}\approx
E^{(k)}<0$ and at $E_{2}\approx E^{(m)}>0$.  Close to these energies,
the phases $\varphi^{(k)}_{1}$ and $\varphi^{(m)}_{2}$, defined in
Eq.~(\ref{resphase}), are close to zero, and we find the current
spectral density for injection of a quasiparticle from the
left (see Appendix~\ref{threepartres}),
\begin{equation}
\label{Ip3res}
\left[J_{3}\right]^{(km),l}=\frac{ D^{3} N^{l}(E)N^{r}(E_{3}) }
{\left|D-4\varphi_{1}^{(k)}\varphi_{2}^{(m)} +
iD\left(\varphi_{1}^{(k)}N^{r}(E_{3})+\varphi_{2}^{(m)}N^{l}(E)
\right)\right|^{2}} .
\end{equation}
We now expand $\varphi_{1}^{(k)}$, $\varphi_{2}^{(m)}$ in the
deviation from perfect overlap in energy, $\delta
E=(E_{1}-E^{(k)}+E_{2}-E^{(m)})/2$, and in voltage, $\delta
V=V-V^{(km)}$, and find, using $D\ll1$, from Eq.~(\ref{Ip3res})
\begin{equation}
\label{doublebreit}
\left[J_{3}(E)\right]^{(km),l} =\frac{D \Gamma_{0}^{(k)}
\Gamma_{3}^{(m)}} {\left(\delta E_{+}\delta E_{-}/\Delta^{2} -
D/4\eta^{(k)} \eta^{(m)}
\right)^{2}+\Lambda^{2}},
\end{equation}
where $\Lambda=(\Gamma_{3}^{(m)}\delta E_{+}
+\Gamma_{0}^{(k)}\delta E_{-})/\Delta$,
$\delta E_{\pm} =\delta E \pm e\delta V/2$.
The energy dependence of the current in Eq. (\ref{doublebreit}) has
the form of two resonant peaks with width $\sim\Gamma\Delta\sim
D\Delta/\eta$ split by the energy interval $\sim\sqrt D\Delta/\eta$ at
$\delta V=0$, the peak splitting increasing with increasing $\delta
V$.  After integration over energy, the overlapping resonances give a
current contribution in the form of a current peak ($k_{B}T\ll\Delta$),
\begin{equation}
\label{I3res}
I^{(km)}_{3}(\delta V)=\frac{3De}{h}\frac{\displaystyle\pi \Delta}{
1+\eta^{(k)}\eta^{(m)}\left(\displaystyle\frac{e\delta
V}{\sqrt{D}\Delta}\right)^{2}}\;
\frac{2N^{l} N^{r}}
{\eta^{(k)}N^{r} +\eta^{(m)} N^{l}}.
\end{equation}
In this equation, the densities of states $N^{l,r}$ are taken at the
entrance and exit energies, $N^{l}\left(2E^{(k)}-E^{^{(m)}}\right)$
and $N^{r}\left(2E^{(m)}-E^{(k)}\right)$, and the temperature is taken
to be zero. A factor of $2$ has been included in Eq.~(\ref{I3res}) to
take into account the similar resonant process for injection from the
right, where $E_{1}=-E^{(m)}$ and $E_{2}=-E^{(k)}$.

The curve for the three-particle current versus voltage thus consists
of peaks with heights of order $D$ and half-width $\Gamma_{V} =
\sqrt{D}\Delta/\eta$ on top of a background of order $D^{2}$. The
background current increases with voltage in the interval
$2\Delta/3<eV<\Delta$ as more single resonances come into the
integration region. In the interval $\Delta<eV$, the background
current decreases due to broadening of the resonances because of
leakage associated with incomplete Andreev reflection outside the gap.

\begin{figure}[ht]
  \begin{center}
    \scalebox{0.32}{\includegraphics{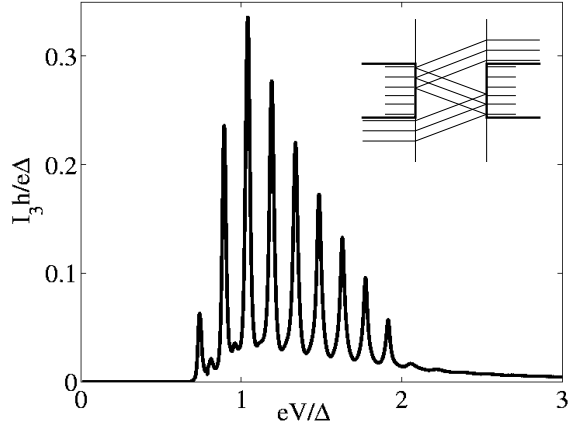}}
  \end{center}
  \caption{``Peak triangle'' of three-particle current for long junction: $L_{l}=L_{r}=L=10\xi_{0}$, $D=0.1$. Every peak of the triangle consists of a number of tightly positioned resonances due to nearly equidistant bound state spectrum (resonance cluster). The number of resonances in a cluster is, from left to right, 1, 4, 7, 6, 5, 4, 3, 2, 1. The inset shows an example of resonant MAR paths forming a cluster. }
  \label{I3long}
\end{figure}

In long symmetric junctions the current peaks form an interesting
triangular pattern. To see this, we first note that if the bound state
spectrum were perfectly linear, several of the peaks described by
Eq.~(\ref{Ip3res})-(\ref{I3res}) will be situated at the same voltage
since $eV^{(km)}=E^{(m)}-E^{(k)}= E^{(m+1)}-E^{(k+1)}$
(see also the inset in Fig.~\ref{I3long}), and thus the total number
of peaks will be reduced while their respective height will be increased.
Since the bound state spectrum is not linear, the peaks show
splitting. However, the deviation from linearity is small and in
practice the peaks form clusters, giving combined peaks with height
roughly equal to the number of clustering peaks.  In the interval
$2\Delta/3<eV<\Delta$, the number of peaks in a cluster increases in
steps of three from $1$ to $4$, etc., up to the number of bound
states. In the interval $\Delta<eV<2\Delta$ the number of peaks in a
cluster decreases in steps of one.  This gives an appearance of a
``peak triangle'' for very long junctions, shown in Fig.~\ref{I3long}.
This ``peak triangle'' is further enhanced by the background
current, which has a similar triangular form, as explained above.

\subsection{Four-particle current}

\begin{figure}[ht]
  \begin{center}
    \scalebox{0.36}{\includegraphics{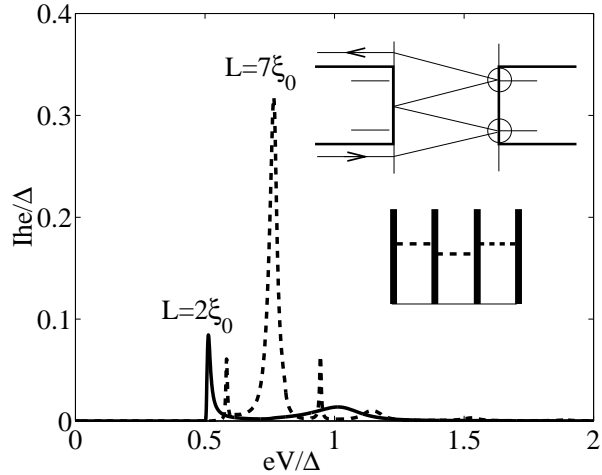}}
  \end{center}
  \caption{Four-particle current in symmetric junctions
    $L_{l}=L_{r}=L/2$ for different lengths; the junction
    transparency is $D=0.1$. The four-particle current in short
    junctions is not visible on the scale in the figure.  The
    solid-line peak and the small dashed-line peaks are due to double
    resonances, illustrated by the MAR diagram in the upper inset.
    Large dashed-line peak is due to a quasi-triple resonance in the MAR
    path. An effective four-barrier structure equivalent to this MAR
    path is shown in the lower inset.}
  \label{I4D1plots}
\end{figure}

The four-particle current has a non-resonant value of order $D^{4}$,
which is enhanced to order $D^{3}$ when the energy of one of the three
Andreev reflections coincides with a bound state energy. Similar to
the three-particle current, overlapping resonances can enhance the
magnitude of the current $I_{4}$ to the order $D$ for those voltages
where both the first and the third Andreev reflections coincide with
the bound states, as shown in the upper inset in Fig.~\ref{I4D1plots}.
Indeed, it is clear from the explicit form of $\hat{M}_{40}=\hat{T}
e^{i\sigma_{z}\varphi_{3}}
\hat{T}^{-1}e^{i\sigma_{z}\varphi_{2}}\hat{T}
e^{i\sigma_{z}\varphi_{1}}\hat{T}^{-1}$ that when $\varphi_{1}=k\pi$
and $\varphi_{3}=m\pi$, then
$\hat{M}_{40}=(-1)^{k+m}e^{i\sigma_{z}\varphi_{2}}$, i.e. the
transparency of the MAR trajectory is enhanced to unity. Other
combinations of the resonances, e.g.  when the first and the second
Andreev reflection occur at bound state energies, will produce peaks
of order $D^{2}$ or smaller, as described in Appendix~\ref{fourpartres}.

Focusing on the double resonances which produce large ($\sim D$)
current peaks, we find that in short junctions with just one pair of
bound states, $\pm E^{(0)}$, the double resonance will occur at
voltage $eV=E^{(0)}$, provided the energy of the bound state is within
the interval $\Delta/2\leq E^{(0)}\leq\Delta$.  The spectral density
of the current has a form similar to the one in
Eq.~(\ref{doublebreit}), the major difference being the small peak
splitting\cite{note2}, $\sim D\Delta/\eta$.
The height of the resulting current peak ($k_{B}T\ll\Delta$) is
\begin{equation}
\label{I4max}
\left[I_{4}\right]_{max}=\frac{\pi De\Delta}{h}
\frac{1-\left[a(2E^{(0)})\right]^{4}}
{1+\left[a(2E^{(0)})\right]^{4}}  ,
\end{equation}
where $a(2E^{(0)})$ is the Andreev reflection amplitude at energy
$2E^{(0)}$.

For longer junctions, there are many possibilities to have overlapping
resonances. Two bound states at one side of the junction with
energies $E^{(k)}<0$ and $E^{(m)}>0$ can give a peak in $I_{4}$ if
$(E^{(m)}-E^{(k)})/2=eV\geq \Delta/2$. Although the height of all peaks
is roughly proportional to $D$, numerically the heights (and widths) of
the peaks may vary considerably depending on the position of the
second Andreev reflection.  If the second Andreev reflection does not
occur at the energy of a bound state, the situation is similar to the
one described above; see lower inset in Fig.~\ref{I4D1plots}. However,
if a bound state is close to the energy of the second Andreev
reflection, then the current spectral density $I^{p}_{4}(E)$ consists
of the three full-transmission peaks with widths $\sim D\Delta/\eta$
which are split within the interval $\sim \sqrt{D}\Delta/\eta$ (triple
resonance). The triple resonance has larger spectral weight compared to
the double resonance, which results in the larger height and width of
the current peak.

Rigorously speaking, a triple resonance can only occur in asymmetric
junctions because it requires equal distance between neighboring
resonances, while the bound state spectrum in symmetric junctions is
not equidistant. However, in long junctions, the deviation from the
equidistant spectrum is small, and quasi-triple resonances may
therefore occur also in long symmetric junctions.
This effect can be observed in Fig.~\ref{I4D1plots}, where the
four-particle current for a symmetric junction with length
$L=7\xi_{0}>2\pi\xi_{0}$ consists of three peaks with different heights:
the central peak corresponding to the quasi-triple resonance while the
two
side peaks corresponding to the double resonances with the heights given
by
Eq.~(\ref{I4max}).

Finally, it is worth noting that, similar to the situation for the
three-particle current, the peaks will form clusters, giving a smaller
number of current peaks than the number of pairs of bound states in
long junctions.

\subsection{High order currents}

The studied properties of multiple resonances in three- and
four-particle currents allow us to make some general conclusions about
resonant behavior of the high order multi-particle currents which
determine the total current at small voltage. The non-resonant magnitude
of an $n$-particle current is of order $D^n$ at the threshold voltage,
$eV_n=2\Delta/n$, and therefore the total non-resonant current
exponentially decreases with the applied voltage (in transparent junctions, $D\sim 1$, the current is exponentially small at\cite{bratus.prb.97} $eV<\Delta\sqrt{1-D}$).  However, multiple resonances may enhance the
magnitude of the current by several orders of $D$. The major question
of interest here concerns the maximum value of the resonant current,
in particular whether it can be of order $D$ at arbitrary small
voltage.

To obtain such large current at small voltage, it is necessary to
achieve a transmission probability through a high order MAR path equal
to unity, which implies that the energy of at least every other
Andreev reflection must coincide with a bound state (cf. the
discussion in the previous subsection).  For $n>4$, this means that
three or more bound states must be approximately equidistant in
energy. Since the bound state spectrum is non-equidistant, Eq.
(\ref{resphase}), this is generally not possible if the resonances are
narrow; therefore, in junctions with arbitrary geometry and small
transmissivity there are no large current peaks below the voltage
$eV=\Delta/2$.

\begin{figure}[ht]
  \begin{center}
    \scalebox{0.4}{\includegraphics{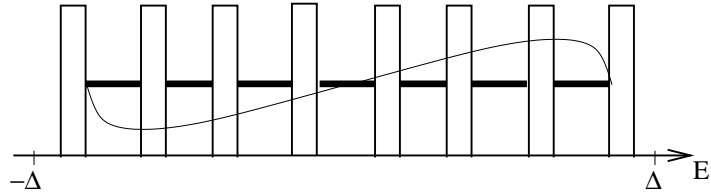}}
  \end{center}
  \caption{ Mapping of a high-order MAR path on a multibarrier
    structure: for an equidistant spectrum, full alignment of
    positions of bound levels (indicated by bold lines) is possible
    for voltage $eV=E^{(m+1)}-E^{(m)}$, yielding a full-transmission
    band. The deviation of the real bound level spectrum from a best
    linear fit is shown by the thin line.  }
  \label{quasi-resonance}
\end{figure}

\begin{figure}[ht]
  \begin{center}
    \scalebox{0.4}{\includegraphics{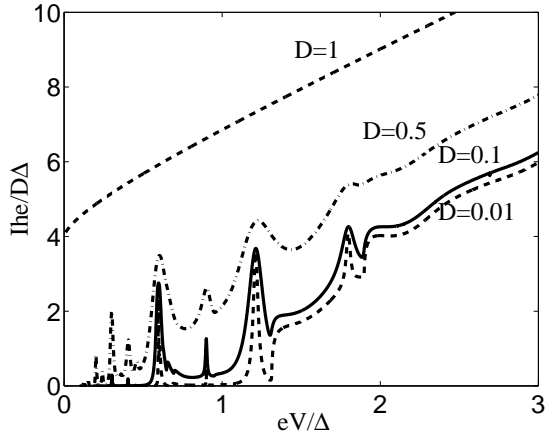}}
  \end{center}
  \caption{Total current for different junction transparencies
    $D$. The junction is symmetric, with length 
    $L_{l}=L_{r}=2\xi_{0}$.}
  \label{ItotL4plots}
\end{figure}

However, the possibility of a large resonant current exists for
junctions with sufficiently large transparency. To find the relevant
transparency, let us consider a very long symmetric junction and
assume for the moment that the bound state spectrum is equidistant, $
E^{(m+1)}-E^{(m)}=const$. Then, from mapping of the $n$-th order MAR
process on a 1D multi-barrier structure (see
Fig.\ref{quasi-resonance}), it is clear that if the applied voltage is
commensurate with the level spacing, e.g., $eV=E^{(m+1)}-E^{(m)}$, the
multi-barrier structure is periodic, and full transmission is achieved
leading to a current peak. This conclusion is valid also for a
non-equidistant spectrum if the variation of the interlevel distance
does not exceed the width of the full-transmission band.  The
deviation of the bound state spectrum from the best linear fit does
not exceed the value 0.33$\Delta\xi_0/L$, Fig.\ref{quasi-resonance}.
On the other hand, the width of the full-transmission energy band is
$\sim\sqrt{D}\Delta/\eta$ for equidistant spectrum and for
$n\rightarrow\infty$. Thus one should expect large current structures
in long symmetric junctions with transparency $D>0.1$ to occur at
voltages $eV>\Delta\xi_0/L$.  In junctions with smaller transparency,
large current structures may appear only at $eV>\Delta/2$, as
explained before; see Fig.~\ref{ItotL4plots}. It is also easy to see
that in asymmetric junctions, where the width of the full transmission
band for an equidistant spectrum is $\sim D\Delta/\eta$ (since the
relevant resonances at one side of the junction are weakly coupled to
each other through the MAR process), large resonant current at small
voltage may exist if $D>0.33$.  Our numerical investigations confirm
that in symmetric junctions when $D$ is of the order $10^{-3}$, the
multiple resonances are completely blocked and current peaks are
exponentially suppressed at $eV<\Delta/2$.

\begin{figure}[ht]
  \begin{center}
    \scalebox{0.4}{\includegraphics{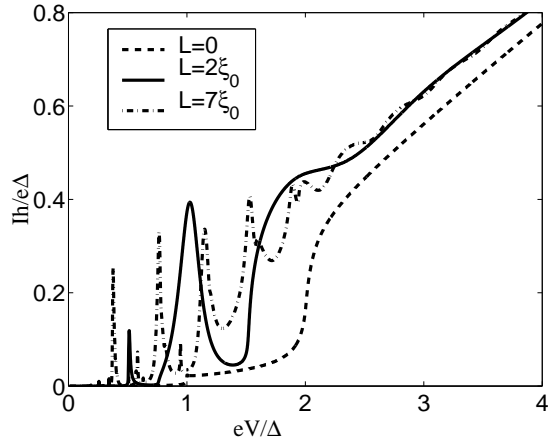}}
  \end{center}
  \caption{Total current in symmetric junctions $L_{l}=L_{r}=L/2$, for
    different lengths; the junction transparency is $D=0.1$.} 
  \label{ItotD1plots}
\end{figure}
\begin{figure}[ht]
  \begin{center}
    \scalebox{0.4}{\includegraphics{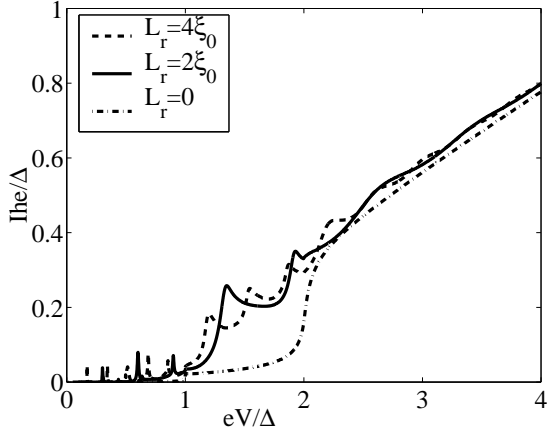}}
  \end{center}
  \caption{Total current in asymmetric junctions $L_{l}=0$, $L_{r}=L$,
    for different lengths; the transparency is $D=0.1$.} 
  \label{ItotD1asymmplots}
\end{figure}

\section{Summary}
\label{discussion}

Adding up the contributions to the current calculated in this paper, we
arrive at a rather complex form of current-voltage characteristics (IVC) at subgap voltages, as shown in Figs.~\ref{ItotD1plots}-\ref{ItotL4plots}.
Nevertheless, the analysis of the tunnel limit allows us to
classify various subgap current structures. Here we will
summarize the results of this classification. As a reference system we
will take a short ($L=0$) junction where the form of the IVC is well
studied \cite{shumeiko.lowtempphys.97}.
The current structures in
short junctions can be interpreted as resonant features due to
quasi-bound states situated at the edges of the energy gap
\cite{johansson.superlatt.99}, the resonant conditions selecting
voltages equal to the gap subharmonics, $eV=2\Delta/n$. This
subharmonic gap structure of the short junction gradually changes with
increasing junction length as bound states move down into the gap,
giving rise to IVC structures with steps, oscillations and peaks. Major points are:

(i) The current in the subgap region is considerably enhanced,
compared to the short junction case. This effect is present as
soon as the effective length $L/\xi_{0}$ is comparable to, or
larger than, the square root of transparency of the junction,
$L/\xi_{0}\sim\sqrt{D}$.

(ii) The main onset of the current in short junctions at $eV=2\Delta$
shifts downwards in voltage to the value $eV=\Delta+E^{(0)}$ where $
E^{(0)}$ is the energy of the bound state. This shift is caused by the
resonant two-particle current giving a contribution to the total
current of the order of the single-particle current.

(iii) For longer junctions, the current onset transforms into a
staircase within the voltage interval $\Delta<eV<2\Delta$ with the
number of steps corresponding to the number of bound states, the step
positions being given by $eV=\Delta+E^{(m)}$. This is due to the
resonances in the two-particle current transported through bound
states.  Resonant channels open up, one by one, as the voltage
increases and bound states enter the ``energy window'' available for
two-particle processes. The current plateaus are not flat but
modulated because of oscillations of the density of continuum states.  The
period of the modulation is roughly equal to the interlevel distance,
and it decreases with the junction length.  The amplitude of the
modulation, on the other hand, increases with the junction length.
Thus, in long junctions, the current structures take the form of a
series of peaks (see Fig.~\ref{ItotD1plots}) within the voltage
interval $\Delta<eV<2\Delta$.  The position of the peaks has
pronounced temperature dependence, scaling with the temperature
dependence of the order parameter, while the distance between peaks has
a weak temperature dependence.

(iv) There is another series of the current peaks whose positions
only weakly depend on temperature and are entirely determined by the
bound state spectrum: $eV= E^{(m)}-E^{(k)}$ and $eV= (E^{(m)}-
E^{(k)})/2$. These peaks are caused by the overlap of two
resonances in the three- and four-particle currents and they exist
in the intervals of applied voltage $2\Delta/3<eV<2\Delta$ and
$\Delta/2<eV<\Delta$ respectively. The heights of these peaks are
comparable with the heights of the two-particle current structures
($\sim D$).

(v) At voltage smaller than $eV=\Delta/2$ the resonant current
structures generally become smaller in magnitude (at least by one
order in $D$) if the junction transparency is sufficiently small
($D\ll 0.1$), and the current decays exponentially when $eV$
approaches zero (although for some particular junction lengths
there could be huge ($\sim D$) current peaks caused by multiple
resonances). This qualitative difference of the IVC below and
above $eV=\Delta/2$ allows one to expect  a cross over from power
to exponential dependence of IVC in multichannel junctions.

(vi) In transparent junctions, all current structures will persist but
become smooth; appreciable current will appear below $eV=\Delta/2$ as
soon as $D\geq 1/3$. The current structures completely disappear in
fully transparent junctions, $D=1$, where the IVC does not depend on
the junction length; see Fig.~\ref{ItotL4plots}.

(vii) At voltage larger than $2\Delta$, the current undergoes
oscillations, similar to Rowell-McMillan oscillations
\cite{rowell.prl.66}, and the excess current is approached much faster
than in short junctions. In low transparency junctions the excess
current is small, $I^{exc}\sim D^2, D\ll1$, despite strong Andreev
reflection and large pair current $I_2\sim D$.

\section*{Acknowledgment}
We thank E.~N. Bratus', J. Lantz and T. L\"ofwander for discussions.
Support from NFR and NUTEK (Sweden) and from NEDO (Japan) is
gratefully acknowledged.

\begin{appendix}
\section{Approximation for \MakeLowercase{$r_{0-}$}
and \MakeLowercase{$r_{n+}$}}
\label{refexp}
In this appendix, the expansion is derived for the reflection amplitudes in 
Eq.~(\ref{reflectionexp}) for a quasiparticle injected from the left.
From the definition of $r_{0-}$ and $r_{(-1)-}$,
Eq.~(\ref{nullvectors}), we know
\begin{equation}
  \hat{ c}_{0-}=c^{\downarrow}_{0-} \left(
    \begin{array}{c}
      r_{0-} \\ 1
    \end{array} \right)
\end{equation}
\begin{equation}
  \hat{ c}_{(-1)-}=c^{\downarrow}_{(-1)-} \left(
    \begin{array}{c}
      r_{(-1)-} \\ 1
    \end{array} \right).
\end{equation}
They are related as
\begin{equation}
  \hat{ c}_{0-}=\hat{M}_{0-1}\hat{U}_{-1}\hat{ c}_{(-1)-},
\end{equation}
where $\hat{M}_{0-1}=\hat{T}$ and
$U_{-1}=e^{i\sigma_{z}\varphi_{-1}}$. From this relation, we
find $r_{0-}$ in terms of $r_{(-1)-}$ as
\begin{equation}
  r_{0-}=\frac{\sqrt{R}+r_{(-1)-}e^{2i\varphi_{-1}}}{1+\sqrt{R}r_{(-1)-
}e^{2i\varphi_{-1}}} = \sqrt{R}
  \left(\frac{1+x/\sqrt{R}}{1-x}\right),
\end{equation}
where $x=(1-\sqrt{R})r_{(-1)-}e^{2i\varphi_{-1}}/(1+r_{(-1)-
}e^{2i\varphi_{-
1}})$. When
$|x|\ll1$, we can make an expansion in this
parameter to get to the form
\begin{equation}
  r_{0-}=\sqrt{R}+
  D\frac{r_{(-1)-}e^{2i\varphi_{-1}}}{1+r_{(-1)-}e^{2i\varphi_{-1}}}
  =\sqrt{R}+O(a_{-1}^{2}D).
\end{equation}
Similarly we also get
\begin{equation}
  r_{n+}=(-1)^{n}\sqrt{R}+
  D\frac{r_{(n+1)+}e^{2i\varphi_{n+1}}}{1+(-1)^{n}r_{(n+1)+}
  e^{2i\varphi_{n+1}}}
  =(-1)^{n}\sqrt{R}+O(a_{n+1}^{2}D).
\end{equation}

\section{Resonance in two-particle current}
\label{twopartres}
In this appendix, we derive the resonant form of the two-particle current,
Eq.~(\ref{ip2res}), for a quasiparticle injected from the left.
The definition of $\hat{M}_{20}$ is
$\hat{M}_{20}=\hat{T}e^{\sigma_{z}\varphi_{1}}\hat{T}^{-1}$,
which using the pseudo-unitarity of the transfer matrices
$\sigma_{z}\hat{T}^{\dagger}\sigma_{z}=\hat{T}^{-1}$ can
be written on the form
\begin{equation}
  \label{M20}
  \hat{M}_{20}=\frac{2i}{\sqrt{D}}
  \sin \varphi_{1}\hat{T}\sigma_{z}+
  e^{-i\sigma_{z}\varphi_{1}} .
\end{equation}
It simplifies in the limit $D\ll1$,
$|\varphi_{1}^{(m)}|=|\varphi_{1}-m\pi|\ll1$ to
\begin{equation}
  \hat{M}_{20}=\frac{(-
1)^{k}}{D}\left(2i\varphi_{1}^{(m)}(1+\sigma_{x})+D\right).
\end{equation}
Inserting the simplified expansion of $\hat{M}_{20}$ and the expansion
of $r_{2+}$ and $r_{0-}$ from Eq.~(\ref{reflectionexp}) into
Eq.~(\ref{Ipn}), as well as putting $R=1$, the leakage current density
takes the form
\begin{equation}
  \left[J_{2}(E)\right]^{l}={\frac{\displaystyle \frac{1-
|a_{0}|^{4}}
      {\left|1-e^{2i\varphi_{0}} \right|^{2}}
      \frac{1-|a_{2}|^{4}}{\left|1-e^{2i\varphi_{2}}
      \right|^{2}} D^{2}}
      {\left|2i\varphi_{1}^{(m)}+\displaystyle\frac{D}{2}\left(
      \displaystyle\frac{1+e^{2i\varphi_{0}}}{1-e^{2i\varphi_{0}}}+
      \displaystyle\frac{1+e^{2i\varphi_{2}}}{1-
e^{2i\varphi_{2}}}\right)
      \right|^{2}}.}
\end{equation}
We make an expansion of the phase
$\varphi_{1}^{(m)}=\eta^{(m)}(E-E^{(m)})/\Delta=\eta^{(m)}\delta
E/\Delta$, where
\begin{equation}
\eta^{(m)}=\Delta\left.\frac{ \partial \varphi}
  {\partial
  E}\right|_{E=E^{(m)}}=\frac{2L_{r}}{\xi_{0}}+\frac{\Delta}
  {\sqrt{\Delta^{2}-(E^{(m)})^{2}}}.
\end{equation}
The two-particle current density now takes a Breit-Wigner form
\begin{equation}
  \left[J_{2}(E)\right]^{l}=\frac{\Gamma_{0}^{(m)}\Gamma_{2}^{(m)}}
  {\displaystyle
  \left(\frac{\delta E-\delta E^{(m)}}{\Delta}\right)^{2}+
  \left(\frac{\Gamma_{0}^{(m)}+\Gamma_{2}^{(m)}}{2} \right)^{2}}
\end{equation}
where the tunneling rates are given by
$\Gamma_{0,2}^{(m)}=N^{l}\left(E_{0,2}\right)D/2\eta^{(m)}$ where
\begin{equation}
  N^{l}\left(E_{0,2}\right)=\mbox{Re}\left\{
    \frac{1+e^{2i\varphi_{0,2}}} {1-e^{2i\varphi_{0,2}}}\right\}
  =\frac{1-|a_{0,2}|^{4}}
  {\left|1-e^{2i\varphi_{0,2}}\right|^{2}}  ,
\end{equation}
are equal to the DOS, Eq.~(\ref{densityofstates}) at energy
$E_{0,2}$. The resonance is slightly shifted from $E^{(m)}$ with
\begin{equation}
  \delta E^{(m)}=\frac{D\Delta}{4\eta^{(m)}}\mbox{Im}
  \left\{\frac{1+e^{2i\varphi_{0}}}{1-e^{2i\varphi_{0}}}+
      \frac{1+e^{2i\varphi_{2}}}{1-e^{2i\varphi_{2}}} \right\}.
\end{equation}

\section{Resonance in three-particle current}
\label{threepartres}
In this appendix, the resonant form of the three-particle current,
Eq.~(\ref{Ip3res}), is derived. The $\hat{M}_{30}$-matrix, which by
definition is
\begin{equation}
  \hat{M}_{30}=\hat{T}^{-1}e^{i\sigma_{z}\varphi_{2}}\hat{T}
  e^{i\sigma_{z}\varphi_{1}}\hat{T}^{-1}  ,
\end{equation}
can be transformed using Eq.~(\ref{M20}) to
\begin{equation}
  \hat{M}_{30}=\hat{T}^{-1}e^{i\sigma_{z}\varphi_{2}}
  \hat{T}\hat{T}^{-1}\hat{T}
  e^{i\sigma_{z}\varphi_{1}}\hat{T}^{-1}
  =\left(\frac{2i}{\sqrt{D}}\sin \varphi_{2}\hat{T}^{-1}\sigma_{z}+
  e^{-i\sigma_{z}\varphi_{2}} \right) \hat{T}^{-1}
  \left( \frac{2i}{\sqrt{D}}\sin \varphi_{1}\hat{T}\sigma_{z}+
  e^{-i\sigma_{z}\varphi_{1}} \right) ,
\end{equation}
which can be written in the form
\begin{equation}
  \label{M30}
  \hat{M}_{30}=-4\sin \varphi_{1}\sin\varphi_{2}
  \left(\hat{1}/\sqrt{D}+\hat{T}^{-1}/D\right) +
  2i\sigma_{z} \sin\left(\varphi_{1}+\varphi_{2}\right) +
  e^{-i\sigma_{z}\varphi_{2}}\hat{T}^{-1}
  e^{-i\sigma_{z}\varphi_{1}} .
\end{equation}

It simplifies in the limit of $D\ll1$,
$|\varphi_{1}^{(k)}|=|\varphi_{1}-k\pi|\ll1$ and
$|\varphi_{2}^{(m)}|=|\varphi_{2}-m\pi|\ll1$ to
\begin{equation}
  \hat{M}_{30}=\frac{(-1)^{k+m}}{D^{3/2}}
  \left[\left(D-4\varphi_{1}^{(k)}\varphi_{2}^{(m)}\right)
  (1-\sigma_{x})+
  Di\sigma_{z}\left(\varphi_{1}^{(k)} (1-\sigma_{x})
  +\varphi_{2}^{(m)}(1+\sigma_{x})\right)\right].
\end{equation}
Inserting this form of the $\hat{M}_{30}$-matrix and the
expansion~(\ref{reflectionexp}) for
$r_{0-}$ and $r_{3+}$ into
Eq.~(\ref{Ipn}), as well as putting $R=1$, the probability
current density for injection of a quasiparticle from the left takes
the form
\begin{equation}
  \left[J_{3}(E)\right]^{l}=\frac{(1-|a_{0}|^{4})(1-
|a_{3}|^{4})D^{3}}
      {\left|1-e^{2i\varphi_{0}} \right|^{2}
      \left|1-e^{2i\varphi_{3}} \right|^{2}\left|Q\right|^{2}}
\end{equation}
\begin{displaymath}
      Q=\left( D-4\varphi_{1}^{(k)}\varphi_{2}^{(m)}
      \right)
      +iD\left(\varphi_{1}^{(k)}
      \frac{1+e^{2i\varphi_{3}}
      }{1-e^{2i\varphi_{3}} }
      +\varphi_{2}^{(m)}\frac{1+e^{2i\varphi_{0}}
      }{1-e^{2i\varphi_{0}} }\right)
\end{displaymath}
where $D\ll1$ is once again used.

Since $|\varphi_{1}^{(k)}|\ll1$ and $|\varphi_{2}^{(m)}|\ll1$ and
the DOS at energies $E_{0,3}$, Eq.~(\ref{densityofstates}), are equal to
\begin{equation}
  N^{l}(E)=\mbox{Re}\left\{
    \frac{1+e^{2i\varphi_{0}}} {1-e^{2i\varphi_{0}}}\right\}
  =\frac{1-|a_{0}|^{4}}
  {\left|1-e^{2i\varphi_{0}}\right|^{2}},
\end{equation}
\begin{equation}
  N^{r}(E_{3})=\mbox{Re}\left\{
    \frac{1+e^{2i\varphi_{3}}} {1-e^{2i\varphi_{3}}}\right\}
  =\frac{1-|a_{3}|^{4}}
  {\left|1-e^{2i\varphi_{3}}\right|^{2}}   ,
\end{equation}
we arrive at the form
\begin{equation}
  \left[J_{3}(E)\right]^{l}=\frac{N^{l}(E)N^{r}(E_{3})D^{3}}
  {\left|D-4\varphi_{1}^{(k)}\varphi_{2}^{(m)} +
  iD\left(\varphi_{1}^{(k)}N^{r}(E_{3})+\varphi_{2}^{(m)}N^{l}(E)
  \right)\right|^{2}} .
\end{equation}

\section{Resonance in four-particle current}
\label{fourpartres}
In this appendix, we discuss the structure of the resonance in the
four-particle current. The matrix
\begin{equation}
  \label{M40}
  \hat{M}_{40}=\hat{T} e^{i\sigma_{z}\varphi_{3}}
  \hat{T}^{-1}e^{i\sigma_{z}\varphi_{2}}\hat{T}
  e^{i\sigma_{z}\varphi_{1}}\hat{T}^{-1},
\end{equation}
can be written as
\begin{equation}
  \label{Mmatris}
  \hat{M}_{40}=\frac{i\sigma_{z}}{D^{2}} \left
    [ -8\sin\varphi_{1} \sin\varphi_{2} \sin\varphi_{3}
    \sqrt{D}\hat{T}^{-1}
    +D\sin\varphi_{1} \sin\varphi_{2}\sin\varphi_{3}
    +D^{2}\sin\left(\varphi_{1}+\varphi_{3}-\varphi_{2} \right)+
  \right.
\end{equation}
\begin{displaymath}
  \left.+
    2D\sin\varphi_{1}\cos\left(\varphi_{3}-
      \varphi_{2}\right) \sqrt{D}\hat{T}^{-1}+
    2D\sin\varphi_{3}\cos\left(\varphi_{1}-
      \varphi_{2}\right) \sqrt{D}\hat{T}^{-1}\right]+
\end{displaymath}
\begin{displaymath}
  +\frac{1}{D^{2}}\left[ -4D\sin\varphi_{1}\sin\varphi_{3}
    \cos\varphi_{2}
    +2D\sin\varphi_{3}
    \sin\left(\varphi_{1}-\varphi_{2}\right)\sqrt{D}\hat{T}+
  \right.
\end{displaymath}
\begin{displaymath}
  \left.+
    2D\sin\varphi_{1}
    \sin\left(\varphi_{3}-\varphi_{2}\right)\sqrt{D}\hat{T}^{-1}+
    D^{2}\cos\left(\varphi_{1}+ \varphi_{3}- \varphi_{2} \right)\right].
\end{displaymath}
From Eq.~(\ref{Mmatris}) it is clear that in general
$\hat{M}_{40}\propto 1/D^{2}$. When both $\varphi_{1}$ and
$\varphi_{3}$ are close to a multiple of $\pi$, $\hat{M}_{40}\propto1$,
while close to other double resonances,
e.g. when $\varphi_{1}$ and $\varphi_{2}$ are close to a multiple of
$\pi$, $\hat{M}_{40}\propto1/D$.

\end{appendix}


\begin{thebibliography}{10}
\expandafter\ifx\csname bibnamefont\endcsname\relax
  \def\bibnamefont#1{#1}\fi
\expandafter\ifx\csname bibfnamefont\endcsname\relax
  \def\bibfnamefont#1{#1}\fi
\expandafter\ifx\csname url\endcsname\relax
  \def\url#1{\texttt{#1}}\fi
\expandafter\ifx\csname urlprefix\endcsname\relax\def\urlprefix{URL }\fi
\providecommand{\bibinfo}[2]{#2}

\bibitem{tomasch.prl.65}
\bibinfo{author}{\bibfnamefont{W.~J.} \bibnamefont{Tomasch}},
  \bibinfo{journal}{Phys. Rev. Lett.} \textbf{\bibinfo{volume}{15}},
  \bibinfo{pages}{672} (\bibinfo{year}{1965}).

\bibitem{rowell.prl.66}
\bibinfo{author}{\bibfnamefont{J.~M.} \bibnamefont{Rowell}}
\bibnamefont{and}
  \bibinfo{author}{\bibfnamefont{W.~L.} \bibnamefont{McMillan}},
  \bibinfo{journal}{Phys. Rev. Lett.} \textbf{\bibinfo{volume}{16}},
  \bibinfo{pages}{453} (\bibinfo{year}{1966}).

\bibitem{rowell.prl.73}
\bibinfo{author}{\bibfnamefont{J.~M.} \bibnamefont{Rowell}},
  \bibinfo{journal}{Phys. Rev. Lett.} \textbf{\bibinfo{volume}{30}},
  \bibinfo{pages}{167} (\bibinfo{year}{1973}).

\bibitem{wolf.85}
\bibinfo{author}{\bibfnamefont{E.~L.} \bibnamefont{Wolf}},
  \emph{\bibinfo{title}{Principles of Electron Tunneling Spectroscopy}}
  (\bibinfo{publisher}{Oxford University Press}, \bibinfo{address}{UK},
  \bibinfo{year}{1985}).


\bibitem{degennes.pl.63}
\bibinfo{author}{\bibfnamefont{P.~G.} \bibnamefont{{de~Gennes}}}
  \bibnamefont{and}
  \bibinfo{author}{\bibfnamefont{D.}~\bibnamefont{{Saint-James}}},
  \bibinfo{journal}{Phys. Lett.} \textbf{\bibinfo{volume}{4}},
  \bibinfo{pages}{151} (\bibinfo{year}{1963}).

\bibitem{morpurgo.prl.97}
\bibinfo{author}{\bibfnamefont{A.~F.} \bibnamefont{Morpurgo}},
  \bibinfo{author}{\bibfnamefont{B.~J.} \bibnamefont{{van Wees}}},
  \bibinfo{author}{\bibfnamefont{T.~M.} \bibnamefont{Klapwijk}},
  \bibnamefont{and}
\bibinfo{author}{\bibfnamefont{G.}~\bibnamefont{Borghs}},
  \bibinfo{journal}{Phys. Rev. Lett.} \textbf{\bibinfo{volume}{79}},
  \bibinfo{pages}{4010} (\bibinfo{year}{1997}).

\bibitem{kastalsky.prl.91}
\bibinfo{author}{\bibfnamefont{A.}~\bibnamefont{Kastalsky}},
  \bibinfo{author}{\bibfnamefont{A.~W.} \bibnamefont{Kleinsasser}},
  \bibinfo{author}{\bibfnamefont{L.~H.} \bibnamefont{Grenee}},
  \bibinfo{author}{\bibfnamefont{R.}~\bibnamefont{Bhat}},
  \bibinfo{author}{\bibfnamefont{F.~P.} \bibnamefont{Milliken}},
  \bibnamefont{and} \bibinfo{author}{\bibfnamefont{J.~P.}
  \bibnamefont{Harbison}}, \bibinfo{journal}{Phys. Rev. Lett.}
  \textbf{\bibinfo{volume}{67}}, \bibinfo{pages}{3026}
(\bibinfo{year}{1991}).

\bibitem{wees.prl.92}
\bibinfo{author}{\bibfnamefont{B.~J.} \bibnamefont{{van Wees}}},
  \bibinfo{author}{\bibfnamefont{P.}~\bibnamefont{{de Vries}}},
  \bibinfo{author}{\bibfnamefont{P.}~\bibnamefont{{Magn\'ee}}},
  \bibnamefont{and} \bibinfo{author}{\bibfnamefont{T.~M.}
  \bibnamefont{Klapwijk}}, \bibinfo{journal}{Phys. Rev. Lett.}
  \textbf{\bibinfo{volume}{69}}, \bibinfo{pages}{510}
(\bibinfo{year}{1992}).

\bibitem{lambert.jphyscondmat.98}
\bibinfo{author}{\bibfnamefont{C.~J.} \bibnamefont{Lambert}}
\bibnamefont{and}
  \bibinfo{author}{\bibfnamefont{R.}~\bibnamefont{Raimondi}},
  \bibinfo{journal}{J. Phys. Cond. Mat.} \textbf{\bibinfo{volume}{10}},
  \bibinfo{pages}{901} (\bibinfo{year}{1998}).

\bibitem{arnold.prb.78}
\bibinfo{author}{\bibfnamefont{G.~B.} \bibnamefont{Arnold}},
  \bibinfo{journal}{Phys. Rev. B} \textbf{\bibinfo{volume}{17}},
  \bibinfo{pages}{3576} (\bibinfo{year}{1978}).

\bibitem{gallaher.prb.80}
\bibinfo{author}{\bibfnamefont{W.~J.} \bibnamefont{Gallaher}},
  \bibinfo{journal}{Phys. Rev. B} \textbf{\bibinfo{volume}{22}},
  \bibinfo{pages}{1233} (\bibinfo{year}{1980}).

\bibitem{blonder.prb.82}
\bibinfo{author}{\bibfnamefont{G.~E.} \bibnamefont{Blonder}},
  \bibinfo{author}{\bibfnamefont{M.}~\bibnamefont{Tinkham}},
\bibnamefont{and}
  \bibinfo{author}{\bibfnamefont{T.~M.} \bibnamefont{Klapwijk}},
  \bibinfo{journal}{Phys. Rev. B} \textbf{\bibinfo{volume}{25}},
  \bibinfo{pages}{4515} (\bibinfo{year}{1982}).

\bibitem{arnold.lowtempphys.85}
\bibinfo{author}{\bibfnamefont{G.~B.} \bibnamefont{Arnold}},
  \bibinfo{journal}{J.~Low Temp. Phys.} \textbf{\bibinfo{volume}{59}},
  \bibinfo{pages}{143} (\bibinfo{year}{1985}).

\bibitem{riedel.prb.93}
\bibinfo{author}{\bibfnamefont{R.~A.} \bibnamefont{Riedel}}
\bibnamefont{and}
  \bibinfo{author}{\bibfnamefont{P.~F.} \bibnamefont{Bagwell}},
  \bibinfo{journal}{Phys. Rev. B}
  \textbf{\bibinfo{volume}{48}},
\bibinfo{pages}{15~198}
  (\bibinfo{year}{1993}).

\bibitem{chaudhuri.prb.95}
\bibinfo{author}{\bibfnamefont{S.}~\bibnamefont{Chaudhuri}}
\bibnamefont{and}
  \bibinfo{author}{\bibfnamefont{P.~F.} \bibnamefont{Bagwell}},
  \bibinfo{journal}{Phys. Rev. B} \textbf{\bibinfo{volume}{51}},
  \bibinfo{pages}{16 936} (\bibinfo{year}{1995}).

\bibitem{klapwijk.physicab.82}
\bibinfo{author}{\bibfnamefont{T.~M.} \bibnamefont{Klapwijk}},
  \bibinfo{author}{\bibfnamefont{G.~E.} \bibnamefont{Blonder}},
  \bibnamefont{and}
\bibinfo{author}{\bibfnamefont{M.}~\bibnamefont{Tinkham}},
  \bibinfo{journal}{Physica B} \textbf{\bibinfo{volume}{109-110}},
  \bibinfo{pages}{1657} (\bibinfo{year}{1982}).

\bibitem{takayanagi.prb.95}
\bibinfo{author}{\bibfnamefont{H.}~\bibnamefont{Takayanagi}},
  \bibinfo{author}{\bibfnamefont{T.}~\bibnamefont{Akazaki}},
\bibnamefont{and}
  \bibinfo{author}{\bibfnamefont{J.}~\bibnamefont{Nitta}},
  \bibinfo{journal}{Phys. Rev. B} \textbf{\bibinfo{volume}{51}},
  \bibinfo{pages}{1374} (\bibinfo{year}{1995}).

\bibitem{chrestin.prb.97}
\bibinfo{author}{\bibfnamefont{A.}~\bibnamefont{Chrestin}},
  \bibinfo{author}{\bibfnamefont{T.}~\bibnamefont{Matsuyama}},
  \bibnamefont{and}
\bibinfo{author}{\bibfnamefont{U.}~\bibnamefont{Merkt}},
  \bibinfo{journal}{Phys. Rev. B} \textbf{\bibinfo{volume}{55}},
  \bibinfo{pages}{8457} (\bibinfo{year}{1997}).

\bibitem{kutchinsky.prl.97}
\bibinfo{author}{\bibfnamefont{J.}~\bibnamefont{Kutchinsky}},
  \bibinfo{author}{\bibfnamefont{R.}~\bibnamefont{Taboryski}},
  \bibinfo{author}{\bibfnamefont{T.}~\bibnamefont{Clausen}},
  \bibinfo{author}{\bibfnamefont{A.}~\bibnamefont{Kristensen}},
  \bibinfo{author}{\bibfnamefont{C.~B.} \bibnamefont{{S\o{}rensen}}},
  \bibinfo{author}{\bibfnamefont{A.}~\bibnamefont{Kristensen}},
  \bibinfo{author}{\bibfnamefont{P.~E.} \bibnamefont{Lindelof}},
  \bibinfo{author}{\bibfnamefont{J.}~\bibnamefont{{Bindslev Hansen}}},
  \bibinfo{author}{\bibfnamefont{C.}~\bibnamefont{{Schelde Jacobsen}}},
  \bibnamefont{and} \bibinfo{author}{\bibfnamefont{J.~L.}
\bibnamefont{Sko}},
  \bibinfo{journal}{Phys. Rev. Lett.}
  \textbf{\bibinfo{volume}{78}},
\bibinfo{pages}{931}
  (\bibinfo{year}{1997}).

\bibitem{bastian.prl.98}
\bibinfo{author}{\bibfnamefont{G.}~\bibnamefont{Bastian}},
  \bibinfo{author}{\bibfnamefont{E.~O.} \bibnamefont{G\"obel}},
  \bibinfo{author}{\bibfnamefont{A.~B.} \bibnamefont{Zorin}},
  \bibinfo{author}{\bibfnamefont{H.}~\bibnamefont{Schulze}},
  \bibinfo{author}{\bibfnamefont{J.}~\bibnamefont{Niemeyer}},
  \bibinfo{author}{\bibfnamefont{T.}~\bibnamefont{Weimann}},
  \bibinfo{author}{\bibfnamefont{M.~R.} \bibnamefont{Bennett}},
  \bibnamefont{and} \bibinfo{author}{\bibfnamefont{K.~E.}
  \bibnamefont{Singer}}, \bibinfo{journal}{Phys. Rev. Lett.}
  \textbf{\bibinfo{volume}{81}},
\bibinfo{pages}{1686}
  (\bibinfo{year}{1998}).

\bibitem{arnold.lowtempphys.87}
\bibinfo{author}{\bibfnamefont{G.~B.} \bibnamefont{Arnold}},
  \bibinfo{journal}{J.~Low Temp. Phys.} \textbf{\bibinfo{volume}{68}},
  \bibinfo{pages}{1} (\bibinfo{year}{1987}).

\bibitem{bratus.prl.95}
\bibinfo{author}{\bibfnamefont{E.~N.} \bibnamefont{Bratus'}},
  \bibinfo{author}{\bibfnamefont{V.~S.} \bibnamefont{Shumeiko}},
  \bibnamefont{and}
\bibinfo{author}{\bibfnamefont{G.}~\bibnamefont{Wendin}},
  \bibinfo{journal}{Phys. Rev. Lett.} \textbf{\bibinfo{volume}{74}},
  \bibinfo{pages}{2110} (\bibinfo{year}{1995}).

\bibitem{averin.prl.95}
\bibinfo{author}{\bibfnamefont{D.}~\bibnamefont{Averin}}
\bibnamefont{and}
  \bibinfo{author}{\bibfnamefont{A.}~\bibnamefont{Bardas}},
  \bibinfo{journal}{Phys. Rev. Lett.} \textbf{\bibinfo{volume}{75}},
  \bibinfo{pages}{1831} (\bibinfo{year}{1995}).

\bibitem{cuevas.prb.96}
\bibinfo{author}{\bibfnamefont{J.~C.} \bibnamefont{Cuevas}},
  \bibinfo{author}{\bibfnamefont{A.}~\bibnamefont{{Mart\'{\i}n-Rodero}}},
  \bibnamefont{and} \bibinfo{author}{\bibfnamefont{A.}~\bibnamefont{{Levy
  Yeyati}}}, \bibinfo{journal}{Phys. Rev. B}
\textbf{\bibinfo{volume}{54}},
  \bibinfo{pages}{7366} (\bibinfo{year}{1996}).

\bibitem{shumeiko.lowtempphys.97}
\bibinfo{author}{\bibfnamefont{V.~S.} \bibnamefont{Shumeiko}},
  \bibinfo{author}{\bibfnamefont{E.~N.} \bibnamefont{Bratus'}},
  \bibnamefont{and}
\bibinfo{author}{\bibfnamefont{G.}~\bibnamefont{Wendin}},
  \bibinfo{journal}{J.~Low Temp. Phys.} \textbf{\bibinfo{volume}{23}},
  \bibinfo{pages}{249} (\bibinfo{year}{1997}).
  
\bibitem{note1} \bibinfo{note}{Although the subharmonic gap structure
    at $eV=2\Delta/n$ in these junctions can be interpreted as
    resonances due to superconducting quasi-bound states situated at
    the gap edges \cite{johansson.superlatt.99}.}

\bibitem{derpost.prl.94}
\bibinfo{author}{\bibfnamefont{N.}~\bibnamefont{{van~der~Post}}},
  \bibinfo{author}{\bibfnamefont{E.~T.} \bibnamefont{Peters}},
  \bibinfo{author}{\bibfnamefont{I.~K.} \bibnamefont{Yanson}},
  \bibnamefont{and} \bibinfo{author}{\bibfnamefont{J.~M.}
\bibnamefont{{van
  Reuitenbeek}}}, \bibinfo{journal}{Phys. Rev. Lett.}
  \textbf{\bibinfo{volume}{73}}, \bibinfo{pages}{2611}
(\bibinfo{year}{1994}).

\bibitem{scheer.prl.97}
\bibinfo{author}{\bibfnamefont{E.}~\bibnamefont{Scheer}},
  \bibinfo{author}{\bibfnamefont{P.}~\bibnamefont{Joyez}},
  \bibinfo{author}{\bibfnamefont{D.}~\bibnamefont{Esteve}},
  \bibinfo{author}{\bibfnamefont{C.}~\bibnamefont{Urbina}},
\bibnamefont{and}
  \bibinfo{author}{\bibfnamefont{M.~H.} \bibnamefont{Devoret}},
  \bibinfo{journal}{Phys. Rev. Lett.} \textbf{\bibinfo{volume}{78}},
  \bibinfo{pages}{3535} (\bibinfo{year}{1997}).

\bibitem{scheer.nature.98}
\bibinfo{author}{\bibfnamefont{E.}~\bibnamefont{Scheer}},
  \bibinfo{author}{\bibfnamefont{N.}~\bibnamefont{Agrait}},
  \bibinfo{author}{\bibfnamefont{J.~C.} \bibnamefont{Cuevas}},
  \bibinfo{author}{\bibfnamefont{A.}~\bibnamefont{{Levi Yeyati}}},
  \bibinfo{author}{\bibfnamefont{B.}~\bibnamefont{Ludoph}},
  \bibinfo{author}{\bibfnamefont{A.}~\bibnamefont{{Mart\'{\i}n-Rodero}}},
  \bibinfo{author}{\bibfnamefont{G.~R.} \bibnamefont{Bollinger}},
  \bibinfo{author}{\bibfnamefont{J.~M.} \bibnamefont{{van~Ruitenbeek}}},
  \bibnamefont{and}
\bibinfo{author}{\bibfnamefont{C.}~\bibnamefont{Urbina}},
  \bibinfo{journal}{Nature} \textbf{\bibinfo{volume}{394}},
  \bibinfo{pages}{154} (\bibinfo{year}{1998}).

\bibitem{ludoph.prb.00}
\bibinfo{author}{\bibfnamefont{B.}~\bibnamefont{Ludoph}},
  \bibinfo{author}{\bibfnamefont{N.}~\bibnamefont{{van der Post}}},
  \bibinfo{author}{\bibfnamefont{E.~N.} \bibnamefont{Bratus'}},
  \bibinfo{author}{\bibfnamefont{E.~V.} \bibnamefont{Bezuglyi}},
  \bibinfo{author}{\bibfnamefont{V.~S.} \bibnamefont{Shumeiko}},
  \bibinfo{author}{\bibfnamefont{G.}~\bibnamefont{Wendin}},
\bibnamefont{and}
  \bibinfo{author}{\bibfnamefont{J.~M.} \bibnamefont{{van Ruitenbeek}}},
  \bibinfo{journal}{Phys. Rev. B} \textbf{\bibinfo{volume}{61}},
  \bibinfo{pages}{8561} (\bibinfo{year}{2000}).

\bibitem{gunsenheimer.prb.94}
\bibinfo{author}{\bibfnamefont{U.}~\bibnamefont{Gunsenheimer}}
  \bibnamefont{and} \bibinfo{author}{\bibfnamefont{A.~D.}
  \bibnamefont{Zaikin}}, \bibinfo{journal}{Phys. Rev. B}
  \textbf{\bibinfo{volume}{50}}, \bibinfo{pages}{6317}
(\bibinfo{year}{1994}).

\bibitem{hurd.prb.96}
\bibinfo{author}{\bibfnamefont{M.}~\bibnamefont{Hurd}},
  \bibinfo{author}{\bibfnamefont{S.}~\bibnamefont{Datta}},
\bibnamefont{and}
  \bibinfo{author}{\bibfnamefont{P.~F.} \bibnamefont{Bagwell}},
  \bibinfo{journal}{Phys. Rev. B}
  \textbf{\bibinfo{volume}{54}},
\bibinfo{pages}{6557}
  (\bibinfo{year}{1996}); {\it ibid \/}
  \textbf{\bibinfo{volume}{56}}, \bibinfo{pages}{11 232}
  (\bibinfo{year}{1997}).

\bibitem{landauer.ibm.57}
\bibinfo{author}{\bibfnamefont{R.}~\bibnamefont{Landauer}},
  \bibinfo{journal}{IBM J.~Res. Dev.} \textbf{\bibinfo{volume}{1}},
  \bibinfo{pages}{223} (\bibinfo{year}{1957}).

\bibitem{buttiker.prl.86}
\bibinfo{author}{\bibfnamefont{M.}~\bibnamefont{{B\"uttiker}}},
  \bibinfo{journal}{Phys. Rev. Lett.} \textbf{\bibinfo{volume}{57}},
  \bibinfo{pages}{1761} (\bibinfo{year}{1986}).

\bibitem{imry.86}
\bibinfo{author}{\bibfnamefont{Y.}~\bibnamefont{Imry}}, in
  \emph{\bibinfo{booktitle}{Directions in Condensed Matter Physics}},
edited by
  \bibinfo{editor}{\bibfnamefont{G.}~\bibnamefont{Grindstein}}
  \bibnamefont{and}
\bibinfo{editor}{\bibfnamefont{G.}~\bibnamefont{Mazenko}}
  (\bibinfo{publisher}{World Scientific}, \bibinfo{address}{Singapore},
  \bibinfo{year}{1986}), p. \bibinfo{pages}{101}.

\bibitem{johansson.physicac.97}
\bibinfo{author}{\bibfnamefont{G.}~\bibnamefont{Johansson}},
  \bibinfo{author}{\bibfnamefont{V.~S.} \bibnamefont{Shumeiko}},
  \bibinfo{author}{\bibfnamefont{E.~N.} \bibnamefont{Bratus'}},
  \bibnamefont{and}
\bibinfo{author}{\bibfnamefont{G.}~\bibnamefont{Wendin}},
  \bibinfo{journal}{Physica C} \textbf{\bibinfo{volume}{293}},
  \bibinfo{pages}{77} (\bibinfo{year}{1997}); 
\bibinfo{author}{\bibfnamefont{G.}~\bibnamefont{Johansson}},
  \bibinfo{author}{\bibfnamefont{E.~N.} \bibnamefont{Bratus'}},
  \bibinfo{author}{\bibfnamefont{V.~S.} \bibnamefont{Shumeiko}},
  \bibnamefont{and}
\bibinfo{author}{\bibfnamefont{G.}~\bibnamefont{Wendin}},
  \bibinfo{journal}{Phys. Rev. B} \textbf{\bibinfo{volume}{60}},
  \bibinfo{pages}{1382} (\bibinfo{year}{1999}).



\bibitem{yeyati.prb.97}
\bibinfo{author}{\bibfnamefont{A.}~\bibnamefont{{Levi Yeyati}}},
  \bibinfo{author}{\bibfnamefont{J.~C.} \bibnamefont{Cuevas}},
  \bibinfo{author}{\bibfnamefont{A.}~\bibnamefont{{Lopez-Davalos}}},
  \bibnamefont{and}
  \bibinfo{author}{\bibfnamefont{A.}~\bibnamefont{{Mart\'{\i}n-Rodero}}},
  \bibinfo{journal}{Phys. Rev. B} \textbf{\bibinfo{volume}{55}},
  \bibinfo{pages}{R6137} (\bibinfo{year}{1997}).


\bibitem{deGennes.book}
\bibinfo{author}{\bibfnamefont{P.~G.} \bibnamefont{{de~Gennes}}},
  \emph{\bibinfo{title}{Superconductivity of Metals and Alloys}}
  (\bibinfo{publisher}{Addison-Wesley}, \bibinfo{address}{Reading,
  Massachusetts}, \bibinfo{year}{1966,1989}).

\bibitem{johansson.superlatt.99}
\bibinfo{author}{\bibfnamefont{G.}~\bibnamefont{Johansson}},
  \bibinfo{author}{\bibfnamefont{G.}~\bibnamefont{Wendin}},
  \bibinfo{author}{\bibfnamefont{E.~N.} \bibnamefont{Bratus'}},
  \bibnamefont{and} \bibinfo{author}{\bibfnamefont{V.~S.}
  \bibnamefont{Shumeiko}}, \bibinfo{journal}{Superlatt. and Microstr.}
  \textbf{\bibinfo{volume}{25}}, \bibinfo{pages}{905}
(\bibinfo{year}{1999}).

\bibitem{bratus.prb.97}
\bibinfo{author}{\bibfnamefont{E.~N.} \bibnamefont{Bratus'}},
  \bibinfo{author}{\bibfnamefont{V.~S.} \bibnamefont{Shumeiko}},
  \bibinfo{author}{\bibfnamefont{E.~V.} \bibnamefont{Bezuglyi}},
  \bibnamefont{and}
\bibinfo{author}{\bibfnamefont{G.}~\bibnamefont{Wendin}},
  \bibinfo{journal}{Phys. Rev. B} \textbf{\bibinfo{volume}{55}},
  \bibinfo{pages}{12 666} (\bibinfo{year}{1997}).

\bibitem{wolfram.68}
\bibinfo{author}{\bibfnamefont{T.}~\bibnamefont{Wolfram}},
  \bibinfo{journal}{Phys. Rev.} \textbf{\bibinfo{volume}{170}},
  \bibinfo{pages}{481} (\bibinfo{year}{1968}).

\bibitem{wendin.superlatt.96}
\bibinfo{author}{\bibfnamefont{G.}~\bibnamefont{Wendin}}
\bibnamefont{and}
  \bibinfo{author}{\bibfnamefont{V.~S.} \bibnamefont{Shumeiko}},
  \bibinfo{journal}{Superlatt. and Microstr.}
\textbf{\bibinfo{volume}{20}},
  \bibinfo{pages}{569} (\bibinfo{year}{1996}).

\bibitem{lofwander.supercond}
\bibinfo{author}{\bibfnamefont{T.}~\bibnamefont{{L\"ofwander}}},
  \bibinfo{author}{\bibfnamefont{V.~S.} \bibnamefont{Shumeiko}},
  \bibnamefont{and}
\bibinfo{author}{\bibfnamefont{G.}~\bibnamefont{Wendin}},
  \emph{\bibinfo{title}{Andreev bound states in high-Tc superconducting
  junctions}}, \bibinfo{note}{review article submitted to Superconductor
  Science and Technology}.

\bibitem{note2} \bibinfo{note}{This results from the fact that the
    resonances are coupled with the MAR trajectory that crosses the
    barrier twice (see upper inset in Fig.~\ref{I4D1plots}), instead
    of once as in the three-particle case, and therefore the resonance
    coupling is weaker. Another difference is that the width of the
    resonance and thus the height of the current peak is independent
    of the DOS at the entrance and exit energies, and only depends on
    the Andreev reflection probability at the exit and entrance
    energies.}

\end{thebibliography}
\end{document}